\documentclass[twocolumn,tighten]{aastex63}
\usepackage{verbatim}
\usepackage{color}
\usepackage{amsmath}
\usepackage{graphicx}
\usepackage{natbib}

\newcommand{\target}{GJ~238}
\newcommand{\planet}{GJ~238~b}

\newcommand{\rsun}{\ensuremath{R_\sun}}
\newcommand{\msun}{\ensuremath{M_\sun}}
\newcommand{\lsun}{\ensuremath{L_\sun}}
\newcommand{\rearth}{\ensuremath{R_\earth}}
\newcommand{\mearth}{\ensuremath{M_\earth}}

\newcommand{\mjup}{\ensuremath{M_{\rm Jup}}}
\newcommand{\teff}{\ensuremath{T_{\rm eff}}}
\newcommand{\teq}{\ensuremath{T_{\rm eq}}}

\newcommand{\feh}{[Fe/H]}

\newcommand{\kms}{\ensuremath{\rm km\,s^{-1}}}
\newcommand{\ms}{\ensuremath{\rm m\,s^{-1}}}
\newcommand{\rstar}{\ensuremath{R_s}}
\newcommand{\mstar}{\ensuremath{M_s}}
\newcommand{\lstar}{\ensuremath{L_s}}

\newcommand{\tess}{{\it TESS}}
\newcommand{\gaia}{{\it Gaia}}

\newcommand{\sig}[1]{\ensuremath{#1\sigma}}
\newcommand{\figr}[1]{Figure~\ref{fig:#1}}
\newcommand{\secr}[1]{Section~\ref{sec:#1}}

\newcommand{\tabr}[1]{\mbox{Table~\ref{tab:#1}}}

\shorttitle{GJ~238~\lowercase{b}}
\shortauthors{Tey et al.}

\begin{document}

\title{\planet: A 0.57 Earth radius planet orbiting an M2.5 dwarf star at 15.2 pc}


\author[0000-0002-5308-8603]{Evan~Tey}
\affiliation{Department of Physics and Kavli Institute for Astrophysics and Space Science, Massachusetts Institute of Technology, 77 Massachusetts Ave, Cambridge, MA 02139, USA}

\author[0000-0002-1836-3120]{Avi Shporer}
\affiliation{Department of Physics and Kavli Institute for Astrophysics and Space Science, Massachusetts Institute of Technology, 77 Massachusetts Ave, Cambridge, MA 02139, USA}

\author{Zifan Lin} 
\affil{Department of Earth, Atmospheric, and Planetary Sciences, Massachusetts Institute of Technology, Cambridge, MA 02139, USA}

\author[0000-0002-3481-9052]{Keivan~G.~Stassun} 
\affiliation{Department of Physics and Astronomy, Vanderbilt University, Nashville, TN 37235, USA}
\affiliation{Department of Physics, Fisk University, Nashville, TN 37208, USA}

\author{Jack~J.~Lissauer}
\affiliation{NASA Ames Research Center, Moffett Field, CA 94035, USA}


\author{Coel Hellier}
\affil{Astrophysics Group, Keele University, Staffordshire, ST5 5BG, UK}




\author[0000-0001-6588-9574]{Karen A.~Collins}
\affil{Center for Astrophysics \textbar \ Harvard \& Smithsonian, 60 Garden Street, Cambridge, MA 02138, USA}

\author[0000-0003-2781-3207]{Kevin I.\ Collins}
\affiliation{George Mason University, 4400 University Drive, Fairfax, VA, 22030 USA}

\author{Geof Wingham}
\affiliation{Mt.~Stuart Observatory, New Zealand}

\author{Howard M. Relles}
\affiliation{Center for Astrophysics \textbar \ Harvard \& Smithsonian, 60 Garden Street, Cambridge, MA 02138, USA}

\author{Franco Mallia}
\affiliation{Campo Catino Astronomical Observatory, Regione Lazio, Guarcino (FR), 03010 Italy}

\author{Giovanni Isopi}
\affiliation{Campo Catino Astronomical Observatory, Regione Lazio, Guarcino (FR), 03010 Italy}

\author[0000-0003-0497-2651]{John F.\ Kielkopf} 
\affiliation{Department of Physics and Astronomy, University of Louisville, Louisville, KY 40292, USA}

\author[0000-0003-2239-0567]{Dennis M.\ Conti}
\affiliation{American Association of Variable Star Observers, 185 Alewife Brook Parkway, Suite 410, Cambridge, MA 02138, USA}

\author[0000-0001-8227-1020]{Richard P.~Schwarz}
\affiliation{Center for Astrophysics \textbar \ Harvard \& Smithsonian, 60 Garden Street, Cambridge, MA 02138, USA}

\author{Aldo Zapparata}
\affiliation{Campo Catino Astronomical Observatory, Regione Lazio, Guarcino (FR), 03010 Italy}

\author[0000-0002-8965-3969]{Steven Giacalone}
\altaffiliation{NSF Astronomy and Astrophysics Postdoctoral Fellow}
\affiliation{Department of Astronomy, California Institute of Technology, Pasadena, CA 91125, USA}



\author{Elise Furlan} 
\affiliation{Caltech/IPAC-NASA Exoplanet Science Institute, 770 S.~Wilson Avenue, Pasadena, CA 91106, USA}


\author{Zachary D.~Hartman}
\affil{Gemini Observatory/NSF’s NOIRLab, 670 A’ohoku Place, Hilo, HI 96720, USA}

\author[0000-0002-2532-2853]{Steve~B.~Howell}
\affil{NASA Ames Research Center, Moffett Field, CA 94035, USA}

\author{Nicholas J.~Scott}
\affil{NASA Ames Research Center, Moffett Field, CA 94035, USA}


\author{Carl Ziegler}
\affiliation{Department of Physics, Engineering and Astronomy, Stephen F. Austin State University, 1936 North St, Nacogdoches, TX 75962, USA}

\author[0000-0001-7124-4094]{C\'{e}sar Brice\~{n}o}
\affiliation{Cerro Tololo Inter-American Observatory/NSF’s NOIRLab, Casilla 603, La Serena 1700000, Chile}

\author{Nicholas Law}
\affiliation{Department of Physics and Astronomy, The University of North Carolina at Chapel Hill, Chapel Hill, NC 27599-3255, USA}

\author[0000-0003-3654-1602]{Andrew W. Mann}
\affiliation{Department of Physics and Astronomy, The University of North Carolina at Chapel Hill, Chapel Hill, NC 27599-3255, USA}


\author[0000-0002-9003-484X]{David~Charbonneau}
\affiliation{Harvard-Smithsonian Center for Astrophysics, 60 Garden St, Cambridge, MA 02138, USA}


\author[0000-0002-2482-0180]{Zahra~Essack}
\affiliation{Department of Physics and Astronomy, The University of New Mexico, 210 Yale Blvd NE, Albuquerque, NM 87106, USA}


\author[0009-0008-5145-0446]{Stephanie Striegel}    
\affiliation{NASA Ames Research Center, Moffett Field, CA 94035, USA}
\affiliation{SETI Institute, Mountain View, CA 94043, USA}




\author{George R.~Ricker} 
\affiliation{Department of Physics and Kavli Institute for Astrophysics and Space Science, Massachusetts Institute of Technology, 77 Massachusetts Ave, Cambridge, MA 02139, USA}


\author{Roland Vanderspek}
\affiliation{Department of Physics and Kavli Institute for Astrophysics and Space Science, Massachusetts Institute of Technology, 77 Massachusetts Ave, Cambridge, MA 02139, USA}

\author[0000-0002-6892-6948]{Sara Seager}
\affiliation{Department of Physics and Kavli Institute for Astrophysics and Space Science, Massachusetts Institute of Technology, 77 Massachusetts Ave, Cambridge, MA 02139, USA}
\affil{Department of Earth, Atmospheric, and Planetary Sciences, Massachusetts Institute of Technology, Cambridge, MA 02139, USA}
\affil{Department of Aeronautics and Astronautics, MIT, 77 Massachusetts Avenue, Cambridge, MA 02139, USA}

\author{Joshua N.~Winn}
\affil{Department of Astrophysical Sciences, Princeton University, Princeton, NJ 08544, USA}

\author{Jon M.~Jenkins} 
\affil{NASA Ames Research Center, Moffett Field, CA 94035, USA},


\correspondingauthor{Avi Shporer}
\email{shporer@mit.edu}

\begin{abstract}
We report the discovery of the transiting planet \planet, with a radius of $0.566\pm0.014$ \rearth\ ($1.064\pm0.026$ times the radius of Mars) and an orbital period of 1.74 day. The transit signal was detected by the \tess\ mission and designated TOI-486.01. The star's position close to the Southern ecliptic pole allows for almost continuous observations by \tess\ when it is observing the Southern sky. The host star is an M2.5 dwarf with $V=11.57\pm0.02$ mag, $K=7.030\pm0.023$ mag, a distance of $15.2156\pm0.0030$~pc, a mass of $0.4193_{-0.0098}^{+0.0095}$ \msun, a radius of $0.4314_{-0.0071}^{+0.0075}$\rsun, and an effective temperature of $3{,}485\pm140$ K. We validate the planet candidate by ruling out or rendering highly unlikely each of the false positive scenarios, based on archival data and ground-based follow-up observations. Validation was facilitated by the host star's small size and high proper motion, of $892.633\pm0.025$ mas per year. 
\end{abstract}


\section{Introduction}
\label{sec:intro}

The search for small transiting planets -- the size of Earth and smaller -- is at the forefront of exoplanet research. Small exoplanets are expected to be composed primarily of silicates and iron, allowing for comparisons to the terrestrial planets of the solar system.
Of course, the transits of small planets are difficult to detect, requiring time series photometry precise enough to detect a relative brightness dip on the order of 10$^{-4}$. In addition, the low masses of small planets are challenging to measure, which is why the mass has not been measured for the majority of known transiting terrestrial planets. Instead, the planets have been ``validated'' through statistical arguments against
the known false-positive scenarios in which the transit-like signal has a non-planetary origin \citep[e.g.,][]{torres11, morton15, morton16}.

The Transiting Exoplanet Survey Satellite \citep[TESS; ][]{ricker14} was designed to detect small transiting planets orbiting bright stars. Especially advantageous stars are nearby M dwarfs with small sizes and high proper motion, properties that facilitate validation \citep[e.g.,][]{vanderspek19, shporer20, gan20, tey23}. 

We present here the discovery of \planet, the smallest planet detected with \tess\
data to this point, with a radius of $0.566\pm0.014$ \rearth\ ($1.064\pm0.026$ Mars radii). The planet orbits an M2.5 dwarf with a proper motion of $892.633\pm0.025$ mas per year, located in the southern \tess\ continuous viewing zone. 
We have analyzed archival data from other telescopes, as well as newly acquired data, that allowed us to conclude that the transit signal seen in the \tess\ data is indeed due to a transiting planet around the M dwarf.

Basic information about \target\ is listed in \tabr{info}. \secr{obs} describes the various data sets we have collected and analyzed. In \secr{star} we provide estimates of the stellar
characteristics of \target\, and in \secr{fps} we describe how we used the available data to reject false positive scenarios. In \secr{dis} we discuss the new planet discovery, and we conclude with a brief summary in \secr{sum}.


\begin{deluxetable}{lcc}
\tablewidth{0pc}
\tabletypesize{\small}
\tablecaption{
    Target information and stellar parameters
    \label{tab:info}
}
\tablehead{
    \multicolumn{1}{c}{Parameter} &
    \multicolumn{1}{c}{Value}    &
    \multicolumn{1}{c}{Source}    
}
\startdata
TIC & 260708537 & TIC V8$^a$\\
R.A. & $\ \ \,$06:33:49.146 & \gaia\ DR3$^b$ \\
Dec. &  -58:31:29.72 & \gaia\ DR3$^b$ \\
$\mu_{RA}$ (mas yr$^{-1}$)  & -404.050 $\pm$ 0.018 & \gaia\ DR3$^b$ \\
$\mu_{Dec}$ (mas yr$^{-1}$) & $ $ 795.950 $\pm$ 0.018 & \gaia\ DR3$^b$ \\
Parallax (mas) &  65.722 $\pm$ 0.013 & \gaia\ DR3$^b$ \\
Distance (pc)  &  15.2156 $\pm$ 0.0030 & \gaia\ DR3$^b$ \\
Epoch & 2016.0 & \gaia\ DR3$^b$\\
$B$ (mag)   & 13.11  $\pm$ 0.03 & UCAC4$^c$ \\
$V$ (mag)   & 11.57  $\pm$ 0.02 & UCAC4$^c$ \\
$R$ (mag)   & 11.22  $\pm$ 0.03 & UCAC4$^c$ \\
$G^d$ (mag) &  10.5328 $\pm$ 0.0028 & \gaia\ DR3$^b$ \\
$T^e$ (mag) &  9.3685 $\pm$ 0.0073  & TIC V8$^a$\\
$J$ (mag)    &  7.898 $\pm$ 0.023 & 2MASS$^f$ \\
$H$ (mag)    &  7.312 $\pm$ 0.034 & 2MASS$^f$ \\
$K$ (mag)    &  7.030 $\pm$ 0.023 & 2MASS$^f$ \\
\hline
\lstar\ (\lsun) &  0.0242$_{-0.0028}^{+0.0032}$ & This work\\ 
\teff\ (K)      & 3,485 $\pm$ 140 & This work\\ 
\feh\           & 0.15 $\pm$ 0.10 & This work\\
\enddata
\tablenotetext{a}{\cite{stassun18a}.}
\tablenotetext{b}{\cite{gaia22}.}
\tablenotetext{c}{\cite{UCAC4}.}
\tablenotetext{d}{\gaia\ band.}
\tablenotetext{e}{\tess\ band.}
\tablenotetext{f}{\cite{cutri03}.}
\end{deluxetable}

\section{Observations and data analysis}
\label{sec:obs}

\subsection{\tess\ data}
\label{sec:tess}

\target\ (TIC 260708537) was observed by \tess\ during the 1st and 3rd years of \tess\ operations, in sectors 1--6 (2018 July 25 -- 2019 January 7), 8--13 (2019 February 2 -- 2019 July 18), and 27--39 (2020 July 4 -- 2021 June 24), for a total of 25 sectors. In the 1st year, also called Cycle 1, it was observed as one of the targets requested by \tess\ Guest Investigator programs G011180 (PI: Courtney Dressing) and G011129 (PI: Wei-Chun Jao). In the 3rd year, also called Cycle 3, it was observed as one of the targets requested by \tess\ Guest Investigator programs G03264 (PI: Jennifer Van Saders), G03278 (PI: Andrew Mayo), and G03272 (PI: Jennifer Burt).

\target\ was initially identified as a transiting planet candidate by the MIT Quick Look Pipeline \citep[QLP,][]{huang20a, huang20b} which processes the \tess\ full frame images (FFIs), using the first six sectors of data. It was designated \tess\ object of interest (TOI) 486.01. Later, it was detected as a transiting candidate by the NASA Science Processing Operations Center (SPOC) pipeline at NASA Ames Research Center \citep{jenkins16}, which analyzes the pre-selected pixel postage stamps recorded at an effective exposure time of 2 minutes.

We have examined several diagnostics of TOI 486.01 before initiating the collection of follow-up data (the latter is described in the following sections). We compared the odd and even transits when folding the \tess\ light curve on twice the candidate orbital period detected by QLP and SPOC, and found that they show consistent depths. We then compared the transit seen when folding the \tess\ light curve on the candidate orbital period to the transit seen when folding the light curve on half the candidate orbital period. We noticed that the latter, at half the candidate orbital period, shows a shallower depth and an increased in-transit scatter. Meaning, the candidate orbital period is the true period. In addition, the SPOC pipeline provides difference images, where data taken in transit is subtracted from data taken out of transit \citep{twicken18}. Analysis of the difference signal position showed it is located $1.65 \pm 3.05$ arcsec from \target.

\begin{figure*}
\includegraphics[width=7.3in]{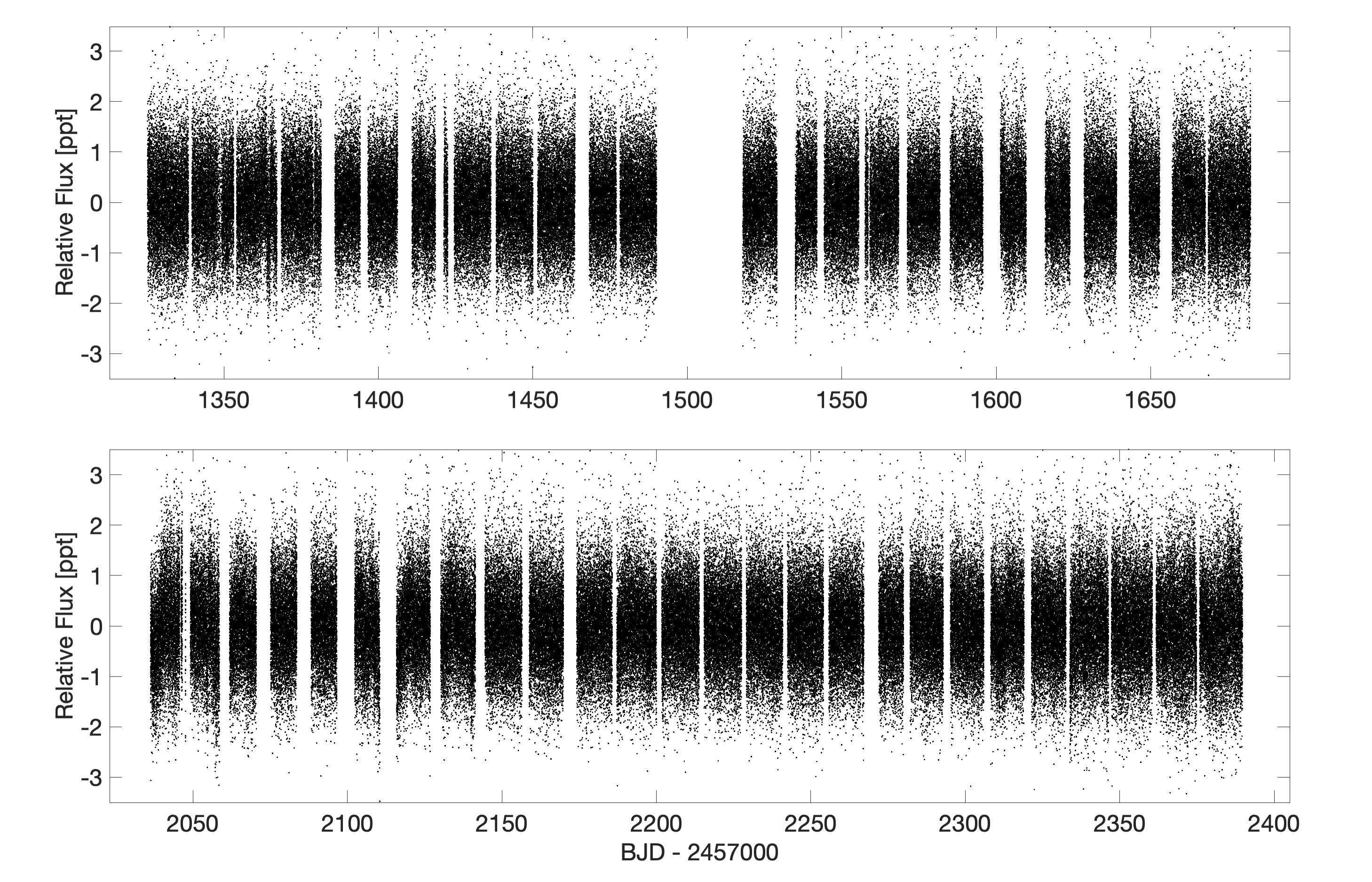}
\caption{\tess\ PDCSAP 2-minute exposure light curve of \target\ plotted in mean-subtracted relative flux (in parts per thousand, or ppt) as a function of time in days. The top panel shows the data in sectors 1--6 and 8--13, and the bottom panel shows data in sectors 27--39. Gaps in the data are primarily due to data downloads and intervals of bad data (due to, e.g., momentum dumps and scattered light). The wide gap in the top panel is during a sector when \target\ was not observed by \tess.
}
\label{fig:tesslc}
\end{figure*}

Since \target\ was observed in 2-minute cadence during all 25 sectors, we chose to use SPOC Pre-search Data Conditioning Simple Aperture Photometry (PDCSAP; \citealt{stumpe12, stumpe14, smith12}) flux for our analysis. The PDCSAP light curve is plotted in \figr{tesslc}.

\subsubsection{\tess\ transit light curve fitting}
\label{sec:trfit}

We fit the \tess\ light curve using \texttt{exoplanet} \citep{foreman21}, which uses PyMC3 to conduct No U-turn sampling over the posterior distribution. In our analysis we ignored data with non-zero quality flags.

For our priors, we used reference parameters from the SPOC Sector 1--39 report. The orbital period and epoch were given uniform priors centered around the values from the SPOC report for sectors 1--39, with a range equal to $\pm 10\%$ of the reference period. The ratio of the planet radius to host star radius $R_p / R_s$ was given a uniform prior from 0 to 2, and the impact parameter $b$ was given a uniform prior from 0 to $(1+R_p/R_s)$. For the stellar mass and radius we used Gaussian priors with means and standard deviations as derived in \secr{star}. Finally, we used a quadratic limb darkening model with parameters $q_1$ and $q_2$ following the priors from \cite{kipping13}.

We sampled five independent chains with 2500 tuning steps and 2500 draws, and confirmed that parameters converged with \cite{gelman92} statistic $\leq 1.01$.

\begin{figure*}
\includegraphics[width=9.0cm]{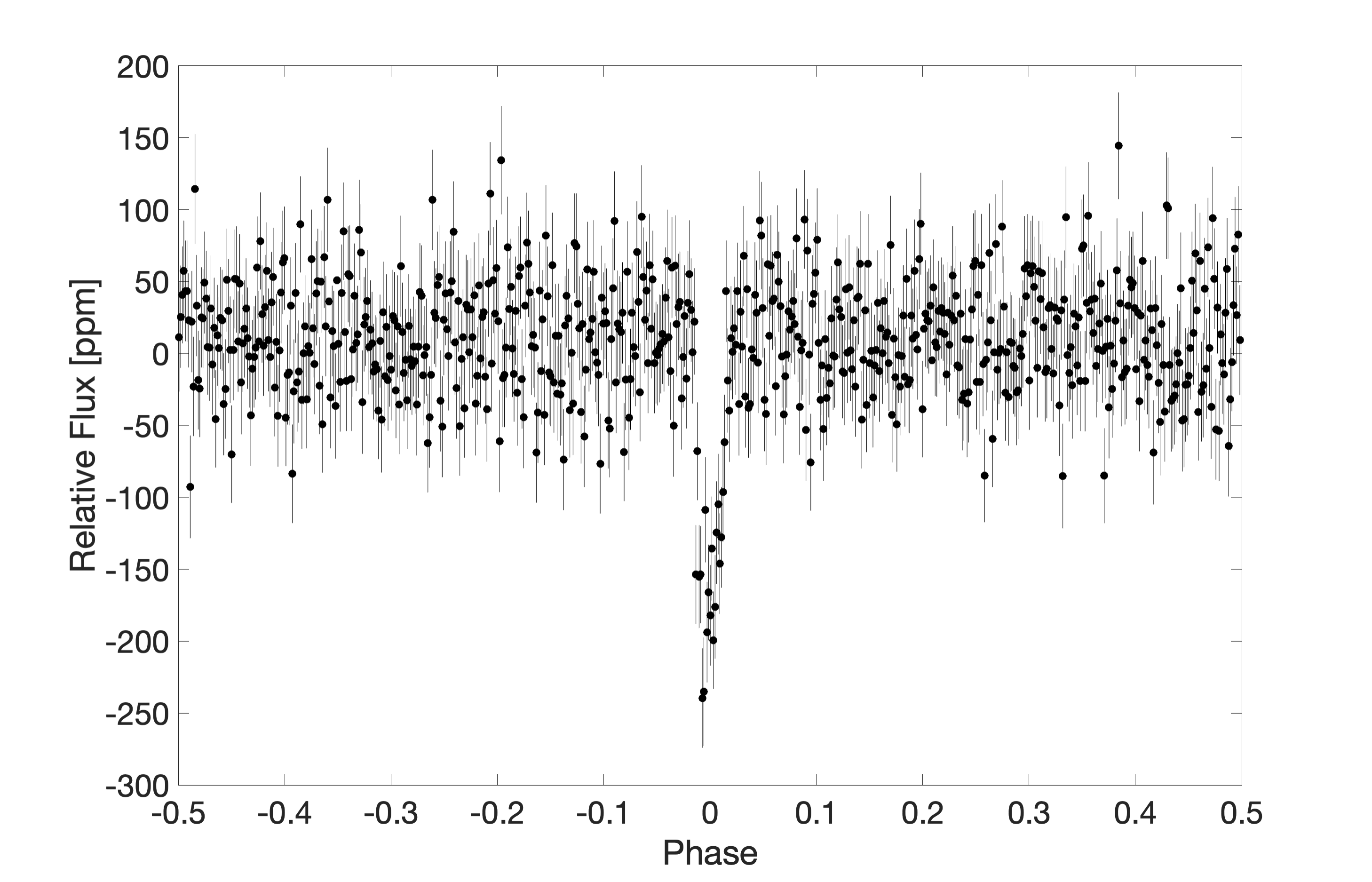}
\includegraphics[width=9.0cm]{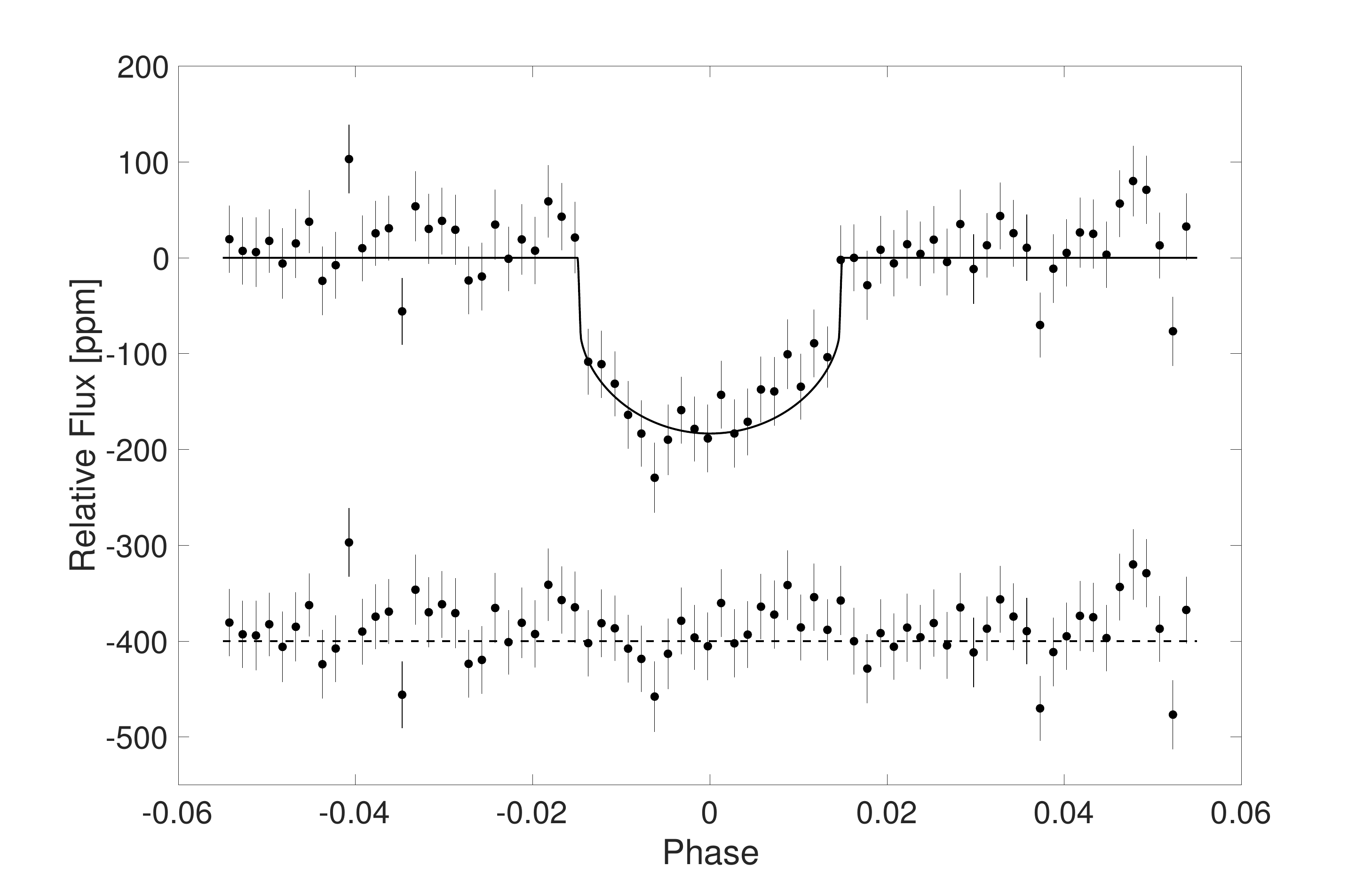}
\caption{Left: \planet\ \tess\ phase-folded and binned transit light curve (in parts per million, ppm) covering an entire orbital period with the transit at phase zero. Right: Phase-folded and binned transit light curve zoomed-in on the transit phase. The best fit model is shown with a black solid line. The lower part of the panel shows the residual flux, after subtracting the model from the data, and offset by 400 ppm, marked by the dashed black line.
}
\label{fig:bestfit}
\end{figure*}

The fitted and derived posteriors are summarized in Table \ref{tab:params}, and the best fit model can be seen in Figure \ref{fig:bestfit}.
While we expect a nearly circular orbit for such a close-orbiting planet (due to tidal dissipation), we conducted an additional fit without this assumption, and we verified that the resulting parameters are consistent with our best fit model
with $e\equiv 0$.
Finally we conducted one last fit in which the stellar mass and radius were given uniform priors between 0 and 1 solar mass and radius, respectively. The results for the stellar mean density were found to be consistent with the mass and radius estimated independently in \secr{star}.

\begin{deluxetable}{lll}
\tablewidth{0pc}
\tabletypesize{\small}
\tablecaption{
    Fitted and derived Parameters
    \label{tab:params}
}
\tablehead{
    \multicolumn{1}{c}{~~~~~~~~~~Parameter~~~~~~~~~~} &
    \multicolumn{1}{l}{Value}                     &
    \multicolumn{1}{l}{Uncertainty}    
}
\startdata
$T_0$ (BJD - 2,457,000) & 1326.98222 & $_{-0.00070}^{+0.00064}$ \\ 
$P$   $\mathrm{(d)}$    & 1.7446999  & $_{-0.0000018}^{+0.0000019}$   \\
$R_p / R_s$             & 0.01204    & $\pm 0.00023$  \\ 
$|b|$                     & 0.202      & $_{-0.072}^{+0.075}$ \\ 
$q_1$ & 0.66 & $\pm 0.16$ \\ 
$q_2$ & -0.04 & $_{-0.17}^{+0.19}$  \\ 
\mstar\ (\msun) & 0.4193 & $_{-0.0098}^{+0.0095}$ \\
\rstar\ (\rsun) & 0.4314 & $_{-0.0071}^{+0.0075}$ \\
$i$ ($^{\circ}$)             & 89.99   & $_{-0.79}^{+0.80}$ \\
$a$ (AU)                     & 0.02123 & $_{-0.00017}^{+0.00016}$ \\
$R_p$ ($\rearth$)              & 0.566   & $\pm 0.014$\\
$T_\mathrm{1,4}$\tablenotemark{a}  (h)    & 1.239 & $\pm 0.018$ \\ 
$q_\mathrm{2,3}$\tablenotemark{b} (h)      & 0.97510 & $_{-0.00109}^{+0.00074}$ \\ 
$q_\mathrm{in}$\tablenotemark{c}  (h) & 0.01245 & $_{-0.00041}^{+0.00054}$ \\ 
$T_\mathrm{eq}$\tablenotemark{d} (K) & 758 &   $_{-15}^{+16}$ \\
\enddata
\tablenotetext{a}{Duration from 1st to last (4th) contacts.}
\tablenotetext{b}{Duration of the total transit, from 2nd to 3rd contacts, as a fraction of the entire transit duration ($T_\mathrm{1,4}$).}
\tablenotetext{c}{Ingress or egress duration as a fraction of the entire transit duration ($T_\mathrm{1,4}$).}
\tablenotetext{d}{Assuming zero Bond albedo and complete heat circulation from the planet's dayside hemisphere (permanently facing the star as the planet is tidally locked) to its nightside hemisphere (permanently facing away from the star).}
\end{deluxetable}

\subsection{Ground-based photometry}
\label{sec:phot}

\subsubsection{Archival imaging}
\label{sec:arch}

\begin{figure*}
\begin{center}
\includegraphics[width=7.1in]{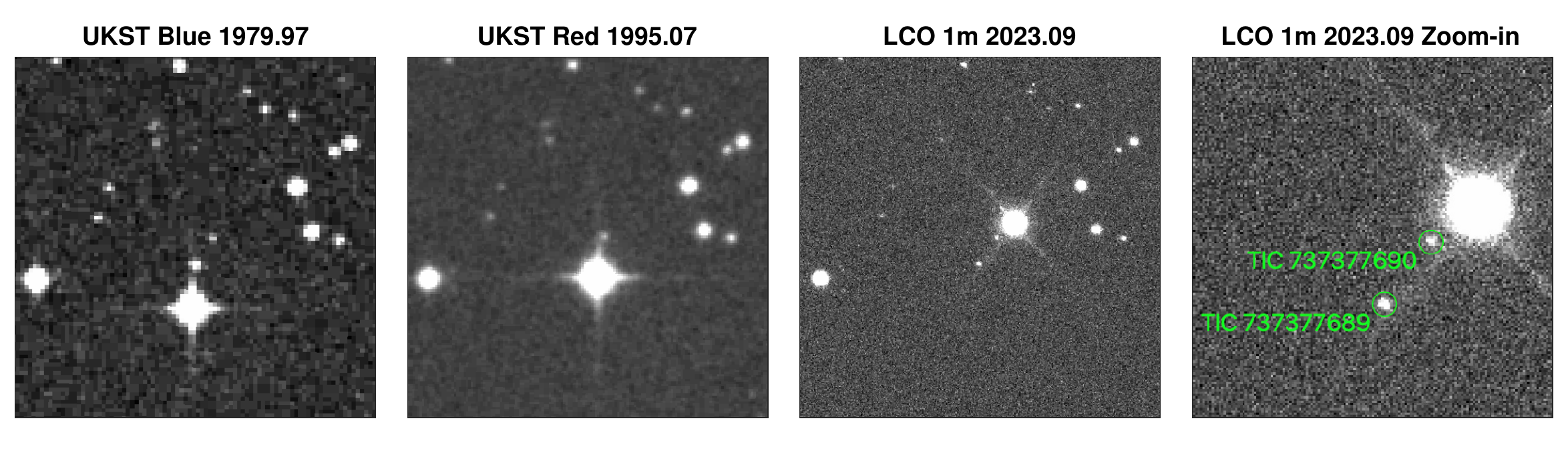}
\end{center}
\vspace{-6mm}
\caption{The two left panels show the target's field of view as observed by the UK Schmidt Telescope (UKST) and processed by the Digitized Sky Survey on 1979 December 21st (leftmost panel) and UT 1995 January 25th (2nd from the left panel). The 3rd panel from the left shows the target's field as observed by an LCO 1~m/Sinistro on UT 2023 February 1st, with the SDSS-$r$ filter. The three left panels show a 2.5 $\times$ 2.5 arcmin field of view centered on the target's position at epoch 2015.5. The motion of the target, which is the brightest star in the field, is clearly seen. The rightmost panel shows a Zoomed-in view of the target and the two stars closest to it on the sky, encircled in green and labeled by their TIC ID. The image is 1.0 $\times$ 1.0 arcmin and the green circles have a 2.0 arcsec radius. In each panel North is up and East is to the left, and the panel's title provides its source and the fractional year when they were obtained.
}
\label{fig:fcharts}
\end{figure*}

We searched for archival images of \target\ and found two exposures taken by the UK Schmidt Telescope (UKST) using photographic plates and made available digitally by the STScI Digitized Sky Survey. One was taken on UT 1979 December 21 with the GG395 filter, otherwise known UKST Blue. The second was taken on UT 1995 January 25 with the RG610 filter, otherwise known as UKST Red. Both exposures are shown in \figr{fcharts} where \target\ is clearly visible. Due to its high proper motion of $892.633\pm0.025$ mas per year \target\ is seen in different locations in the sky in the two archival exposures. It also allows to examine its current location and look for background stars at its position at the time it was observed by \tess\ and ground-based observatories (described in the subsections below), which we discuss in detail in \secr{beb}.

\subsubsection{Las Cumbres Observatory}
\label{sec:lco}

We obtained several ground-based photometric time series covering predicted transit times at several observatories\footnote{Detailed results of each observation is available on ExoFOP-TESS}, including the Las Cumbres Observatory Global Telescope (LCOGT; \citealt{brown13}) 1m telescopes (on UT 2019 March 3, UT 2019 April 7, UT 2019 December 12, UT 2022 November 11, UT 2022 November 18), LCO 0.4m (UT 2019 April 7), Mt.~Kent Observatory (UT 2019 April 7), and Mt.~Stuart 12.5 inch (UT 2019 April 7). We used the {\tt TESS Transit Finder}, which is a customized version of the {\tt Tapir} software package \citep{jensen13}, to schedule our transit observations. While the transit is too shallow to be detected in ground-based data obtained by small telescopes, we obtained these data to look for deep eclipses of nearby stars. Such eclipses, if they exist, might be responsible for the shallow transit-like signals seen in the \tess\ data due to blending with the target. 
Differential photometric data were extracted from the image sequences using {\tt AstroImageJ} \citep{collins17}.

\begin{figure}
\hspace{-4mm}
\includegraphics[width=3.75in]{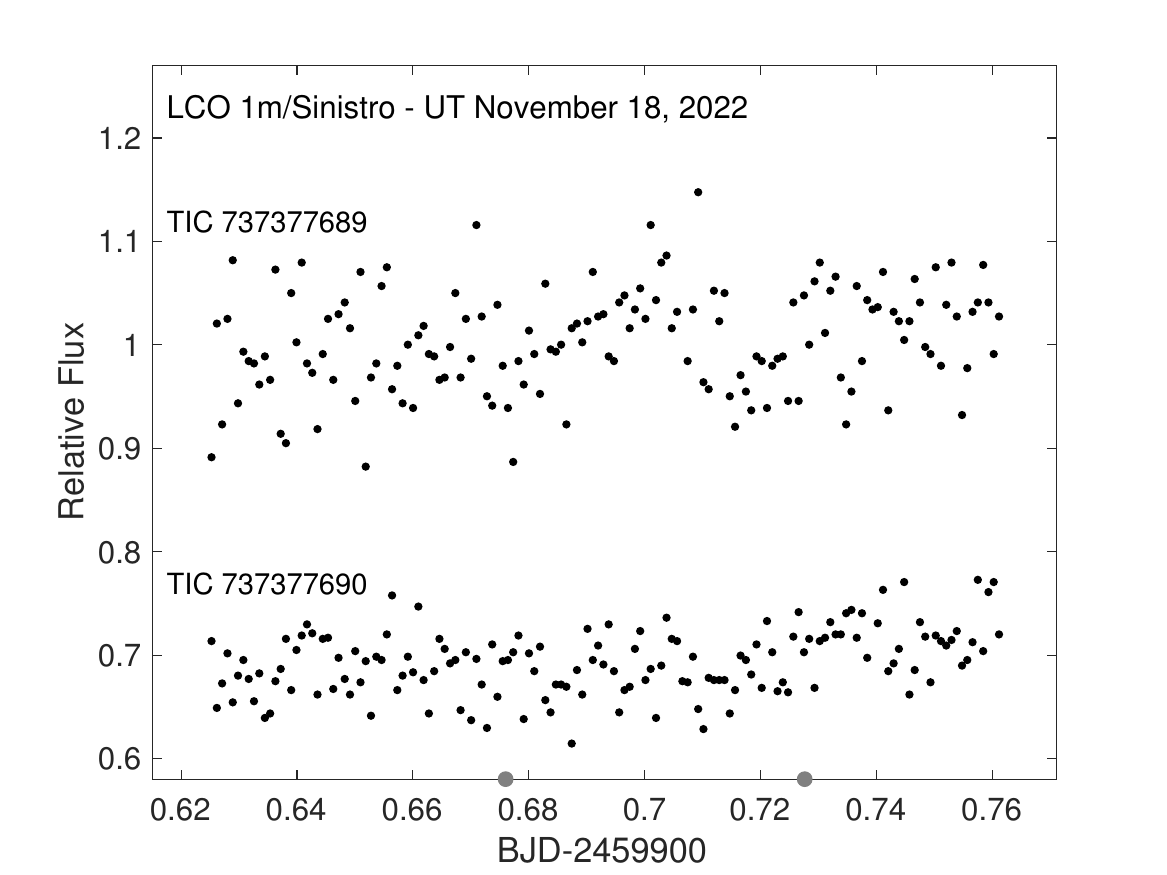}
\caption{Light curves of the two closest stars to the target spanning a predicted transit time, based on observations with an LCO 1m/Sinistro on UT 2022 November 18, through a Pan-STARRS z$_{\rm s}$ filter (overlapping the \tess\ band). Both light curves are plotted in relative flux, and shifted from each other along the Y-axis for clarity. The seeing during this observation was 2.6 arcsec and a 4.0 arcsec aperture diameter was used to extract the light curves (for comparison, the distance between TIC 737377690 and the target was close to 10 arcsec at the time of observation). The two grey points along the X-axis mark the predicted transit start and end times. The uncertainty on those values is similar to the size of the markers. The brightness difference between the target and TIC 737377690 and TIC 737377689 is about 8.2 mag in the \tess\ band and 7.9 mag in $G$ (the two nearby stars have similar brightness and color). Given these brightness differences, to explain the transit seen in \tess\ data one of the light curves would have to show a 25\% deep eclipse. Such deep eclipses can be ruled out.
}
\label{fig:neb}
\end{figure}

Using the data sets listed above, we were able to rule out eclipses deep enough to account for the \tess\ signal, for all the nearby stars. In \figr{neb} we show the results of an observation done by an LCO 1m/Sinistro on UT 2022 November 18, where we could rule out a deep eclipse on the two closest stars to the target, TIC~737377689 ($T$ = 17.62 mag, $G$ = 18.37 mag) and TIC~737377690 ($T$ = 17.61 mag, $G$ = 18.48 mag), marked in \figr{fcharts} rightmost panel. We used the Pan-STARRS z$_{\rm s}$ filter in that observation, as it covers a wavelength range fully within that of the \tess\ band. In that observation TIC~737377689 and TIC~737377690 were well resolved from the target, and the scatter in their light curves (5.0\% and 3.1\%, respectively) completely rules out an eclipse depth of about 25\%, which could have resulted in the transit depth seen in \tess\ data due to blending with the target in \tess\ point spread function (PSF).

\subsubsection{WASP}
\label{sec:wasp}

\begin{figure}
\hspace{-3mm}
\includegraphics[width=3.4in]{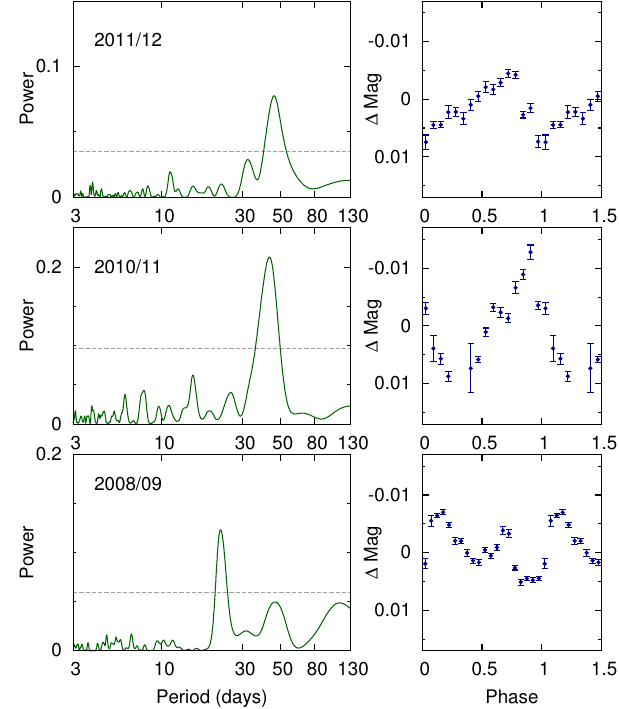}
\caption{Period analysis of WASP times-series photometry. The panels show periodograms of three seasonal data sets (left column panels) and the phase-folded and binned light curve using the putative 44-d rotational period (right column panels). The strongest peak of the 2010/11 and 2011/12 seasons periodogram is at 44 days, while for 2008/09 it is at the 22-day first harmonic. The horizontal dashed lines are the estimated 1\%\ false-alarm levels.
}
\label{fig:wasp}
\end{figure}

We used photometric data collected by WASP-South \citep{pollacco06} during 3 observing seasons, 2008--2009, 2010--2011, and 2011--2012, to try measuring the stellar rotation period using the methods discussed by \citet{maxted11}. The period analysis of each observing season is shown in \figr{wasp}. The strongest period component in the latter two seasons is at $44\pm2$ days, while in the first season (2008--2009) it is close to half that period, at $22\pm1$ days. We adopt $44\pm2$ days as an estimate of the star's rotation period since it is common for the photometric modulation to lose coherence across several years, and it is common for the photometric period to occasionally be one-half of the rotation period due to spots on opposite hemispheres \citep[e.g.,][]{Mcquillan14}.

\subsection{High resolution spectroscopy}
\label{sec:spec}

\begin{figure*}
\includegraphics[width=7.2in]{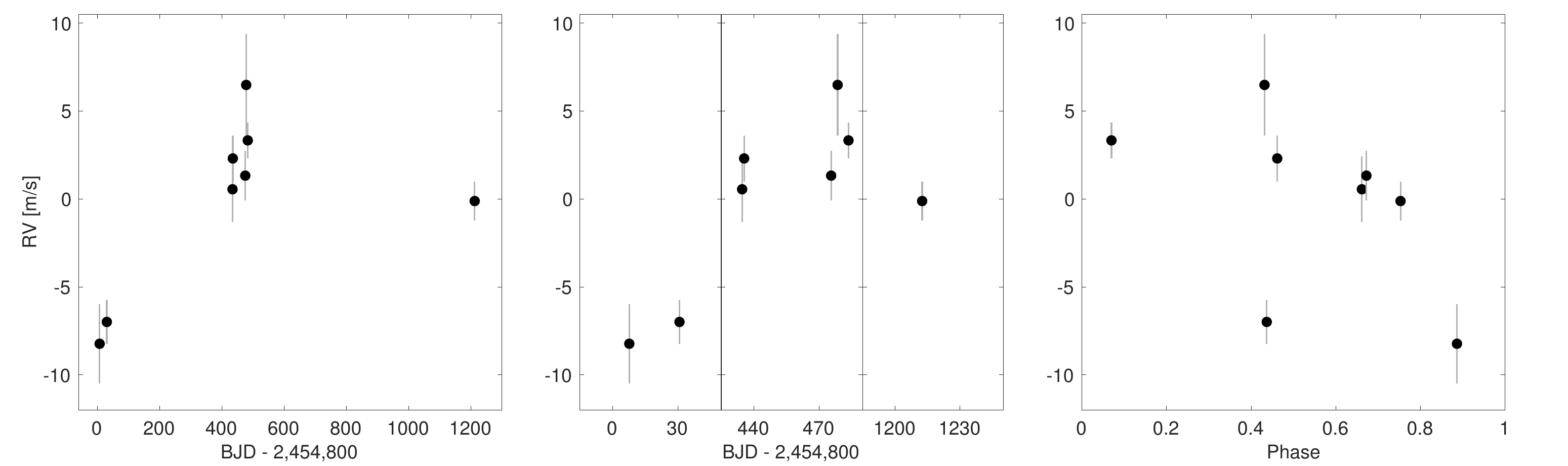}
\caption{Archival HARPS radial velocities (RVs) of \target, from \cite{trifonov20}. The left panel shows the RVs as a function of time, in BJD. The center panel also shows the RVs as a function of time in BJD but while zooming in on the groups of RVs obtained during the same observing season. The right panel shows the RVs phase folded on the transit ephemeris.}
\label{fig:rvs}
\end{figure*}

\target\ was observed by the High Accuracy Radial velocity Planet Searcher \citep[HARPS, $R$=120{,}000;][]{pepe02,mayor03} at the ESO 3.6m telescope in La Silla, Chile, between December 2008 to March 2012. A total of eight spectra were obtained, with a signal-to-noise ratio at order 55 ranging from 15 to 41, under Program IDs 082.C-0718 (PI: Bonfils) and 183.C-0437 (PI: Bonfils). We use radial velocities (RVs) and uncertainties derived from the HARPS spectra by \citet{trifonov20}, who analyzed publicly available HARPS spectra with the public SpEctrum Radial Velocity AnaLyser (SERVAL) pipeline \citep{zechmeister18} and corrected for nightly RV zero points. The 8 RVs, with RV uncertainties ranging from 1.0 to 2.9 \ms, are plotted in \figr{rvs} and listed in \tabr{rvs}, including the cross correlation function FWHM, contrast, bisector span, and signal-to-noise ratio at order 55.

\begin{deluxetable*}{lrrrrrr}
\tablewidth{0pc}
\tabletypesize{\small}
\tablecaption{
    HARPS archival differential radial velocities\tablenotemark{a}
    \label{tab:rvs}
}
\tablehead{
    \multicolumn{1}{c}{BJD} &
    \multicolumn{1}{c}{RV}    &
    \multicolumn{1}{c}{$\sigma_{RV}$}  &
    \multicolumn{1}{c}{FWHM}    &
    \multicolumn{1}{c}{Contrast} &
    \multicolumn{1}{c}{BIS} &
    \multicolumn{1}{c}{SNR}    \\
    \multicolumn{1}{c}{} &
    \multicolumn{1}{c}{\ms}    &
    \multicolumn{1}{c}{\ms}  &
    \multicolumn{1}{c}{\kms} &
    \multicolumn{1}{c}{\ms} &
    \multicolumn{1}{c}{\ms} &
    \\  
}
\startdata
2454807.70656 & -8.23 & 2.26 & 3.058 & 23.15 & -10.8 & 19.5\\
2454830.79427 & -6.99 & 1.25 & 3.046 & 23.17 & -6.2  & 34.3\\
2455234.61573 & 0.55  & 1.86 & 3.045 & 23.03 &  3.2  & 22.6\\ 
2455235.54858 & 2.30  & 1.31 & 3.056 & 23.17 & -2.8  & 31.8\\ 
2455275.56181 & 1.33  & 1.42 & 3.045 & 23.27 & -3.3  & 32.3\\	
2455278.54649 & 6.48  & 2.89 & 3.040 & 23.41 & -22.9 & 14.7\\	
2455283.54149 & 3.33  & 1.01 & 3.045 & 23.22 & -7.8  & 40.8\\	
2456012.53361 &-0.12  & 1.12 & 3.042 & 23.03 & -10.4 & 39.7\\
\enddata
\tablenotetext{a}{RVs and RV uncertainties ($\sigma_{RV}$) are taken from \cite{trifonov20}. Cross correlation function FWHM, Contrast, bisector span (BIS) and signal-to-noise at order 55 (SNR) are taken from the HARPS DRS pipeline.}
\end{deluxetable*}

\subsection{High angular resolution imaging}
\label{sec:imaging}

We obtained high angular resolution images of the target to look for stars located close enough to the target to have been unresolved in seeing-limited ground-based images. In particular, the high angular resolution images can probe for other stars within the HARPS fiber angular radius (see \secr{spec}). If such nearby stars exist they would reduce the relative brightness dip observed with \tess, leading to a smaller planet radius estimate \citep{ciardi15}.
Deep eclipses of nearby faint stars could
also be entirely responsible for the \tess\ signal.

\begin{figure}
\includegraphics[width=\linewidth]{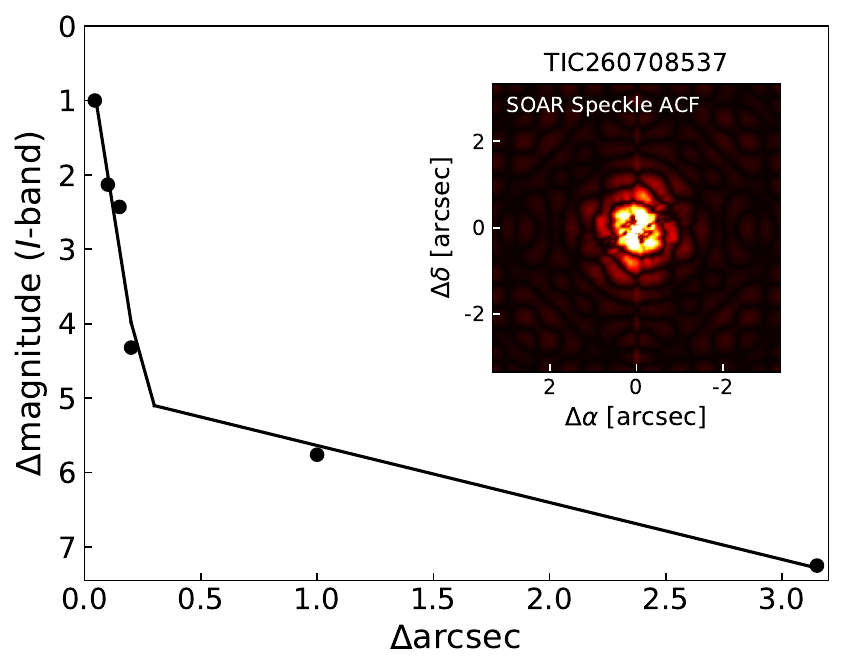}
\includegraphics[width=\linewidth]{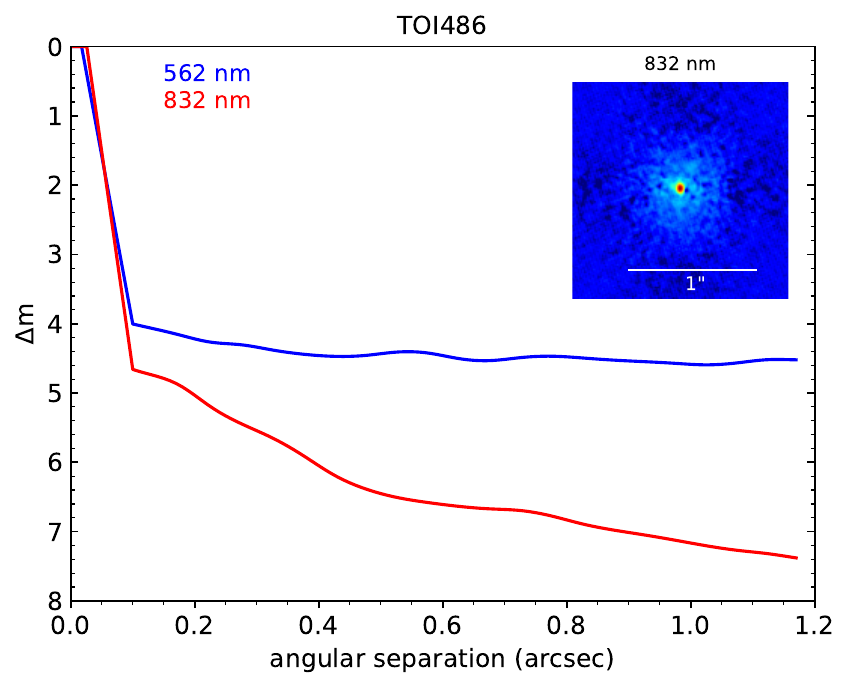}
\caption{Speckle imaging of \target. Each panel shows the contrast curve, in magnitude difference as a function of angular distance from the target in arcsec, and the speckle auto-correlation function (ACF) in the inset. Top: Speckle imaging obtained at SOAR in the $I$ band, on UT 2020 February 10. Bottom: Speckle imaging obtained with Gemini-South/Zorro, on UT 2020 March 13, in 562 nm (blue line) and 832 nm (red line and inset).
}
\label{fig:highangimage}
\end{figure}

We obtained speckle imaging in $I$ band with the high-resolution camera \citep[HRCam,][]{tokovinin18} at the 4.1m Southern Astrophysical Research (SOAR) telescope at Cerro Pach\'on, Chile, on UT 2020 February 10. This observation was done as part of the SOAR \tess\ survey \citep{ziegler21}. No nearby stars were detected within the sensitivity of the data, of $\Delta I$ = 5 mag at 0.5 arcsec and $\Delta I$ = 7 at 3 arcsec, as shown in \figr{highangimage}.

We also obtained speckle imaging with the Zorro speckle instrument \citep{scott21} on the Gemini-South 8m telescope at Cerro Pach\'on, Chile, on UT 2020 March 13. The Zorro instrument observes simultaneously in two bands, 562 nm and 832 nm. No nearby stars were detected within the sensitivity of the data, down to 4 magnitudes fainter than the target at 0.1 arcsec from the target in 562 nm, and in 832 nm down to 4.5 magnitudes at 0.1 arcsec and 7 magnitudes at 1 arcsec, as shown in \figr{highangimage}.

\section{Host star characterization}
\label{sec:star}

We estimated the host star parameters using the empirical relations of \cite{mann15, mann19}, as described below. We first derived the absolute magnitude in $K$ band based on the observed magnitude and \gaia\ DR3 parallax, giving $M_K$ = 6.119 $\pm$ 0.023 mag.
Next, we used the empirical relation between $M_K$ and stellar mass with the coefficients provided by \citet[][see their Table 6 and Equation 2]{mann19}. This resulted in $M_s = 0.421 \pm 0.021\ \msun$, adopting a conservative uncertainty of 5\% \citep{tayar22}. For comparison, using the empirical relation of \citet[see their Table 1 and Equation 10]{mann15} results in $0.444 \pm 0.022\ \msun$, which is 5\% higher than the mass estimate above. 

To estimate the stellar radius we used the empirical relation between $M_K$ and stellar radius derived by \citet[][see their Table 1]{mann15}, resulting in $R_s = 0.427 \pm 0.021\ \rsun$, assuming again a conservative uncertainty of 5\%. For comparison, using the radius-mass empirical relation derived by \citet[][their Equation 10]{boyajian12} and the mass derived in the previous paragraph, results in $0.403 \pm 0.020\ \rsun$, which is 6\% lower than the radius estimate above. 

The preceding estimates of the stellar mass and radius were used as priors during the transit light curve model fitting, described in \secr{trfit}, which led to refined values for the stellar mass and radius, listed in \tabr{params}.  

To estimate the stellar effective temperature we first calculated the bolometric correction BC$_{K}$ to $M_K$. We did that using the empirical relations between BC$_{K}$ and the $V-J$ color given by \citet[][Table 3]{mann15}. We found BC$_{K}$ = 2.66 $\pm$ 0.13 mag and in turn a bolometric magnitude of $M_{\rm bol}$ = 8.78 $\pm$ 0.13 mag, which is equivalent to a bolometric luminosity of \lstar\ = $0.0242_{-0.0028}^{+0.0032}$ \lsun. Finally, using the Stefan-Boltzmann law we found \teff = $3485_{-135}^{+142}$ K. That value is consistent with the temperature estimated by \cite{kuznetsov19}, based on the HARPS spectra. The stellar parameters derived here correspond to a spectral type of M2.5 \citep{pecaut13}\footnote{See also: \url{http://www.pas.rochester.edu/~emamajek/\\EEM\_dwarf\_UBVIJHK\_colors\_Teff.txt}}.

We note that our derived stellar mass, radius, and temperature, are well within \sig{1} from the values in TIC V8 \citep{stassun18a}. And while they are also consistent with the temperature and radius reported by \cite{muirhead18}, the latter reports a stellar mass of 0.473 $\pm$ 0.004 \msun, significantly larger than measured here.  



To estimate the stellar metallicity we used the method described in \cite{dittmann16}. We first identified stars in the sample listed in \cite{dittmann16} that have similar color and $M_K$ to \target, and then calculated a weighted mean of those stars' metallicities, with weights based on the distance from the target position in the color-magnitude parameter space. This led to a metallicity of \feh\ = 0.15~$\pm$~0.10. We note that that value is \sig{1.5} larger than the metallicity reported by \cite{kuznetsov19}, of $-0.05\pm0.09$, based on the HARPS spectra.

\begin{figure}[!ht]
\hspace{-3mm}
\includegraphics[width=1.05\linewidth,trim=80 70 50 50,clip]{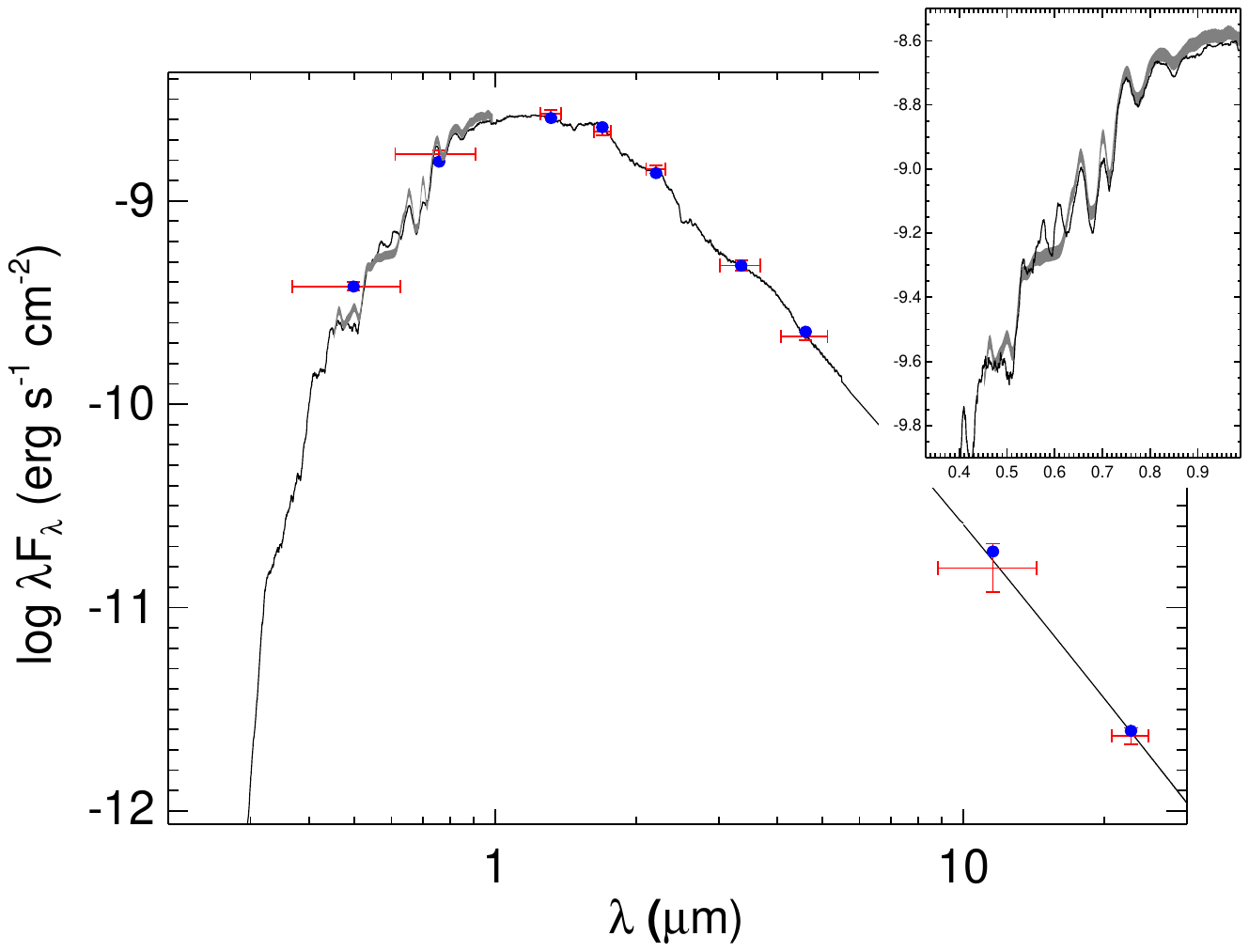}
\caption{\target\ spectral energy distribution. The red symbols represent the observed photometric measurements, and the horizontal bars represent the effective width of the passbands. Blue symbols are the model fluxes from the best-fit PHOENIX atmosphere model (black). The \gaia\ spectrophotometry is shown in gray in both the plot and the inset close-up of the blue end. \label{fig:sed}}
\end{figure}

As an alternative estimate of the stellar temperature and radius we performed an analysis of the entire broadband spectral energy distribution (SED) together with the \gaia\ DR3 parallax following the procedures described in \citet{stassun16,stassun17,stassun18b}. We use the {\it 2MASS} $JHK_S$ magnitudes, the {\it WISE} W1--W4 magnitudes, and the \gaia\ $G_{\rm BP}$ and $G_{\rm RP}$ magnitudes. We also used the \gaia\ spectrophotometry spanning 0.4--1.0~$\mu$m. Altogether, the available photometry spans the full stellar SED over the wavelength range 0.4--20~$\mu$m, with the \gaia\ spectrophotometry providing an especially strong constraint on the overall absolute flux calibration, as shown in \figr{sed}.

We performed a fit using PHOENIX stellar atmosphere models \citep{husser13} and the metallicity estimated above, with the free parameters being the effective temperature and the overall flux normalization. We adopted an extinction $A_V \equiv 0$. The resulting fit (Figure~\ref{fig:sed}) has a reduced $\chi^2 = 1.2$, with $\teff = 3440 \pm 75$~K. Integrating the model SED gives the bolometric flux at the Sun, $F_{\rm bol} = 3.29 \pm 0.12 \times 10^{-9}$ erg~s$^{-1}$~cm$^{-2}$. Taking the $F_{\rm bol}$ and the \gaia\ parallax together directly yields the bolometric luminosity, $L_{\rm bol} = 0.02374 \pm 0.00042$~L$_\odot$, which from the Stefan-Boltzmann relation gives the stellar radius to be $R_s = 0.434 \pm 0.020$~R$_\odot$. The results of the SED analysis provide consistent values for \teff\ and $R_s$ with those provided by the empirical relations.

The HARPS spectra show absorption in the $H_{\alpha}$ line. According to \cite{walkowicz09}, stars of spectral type similar to \target\ that show absorption in $H_{\alpha}$ are either moderately active or inactive, but not strongly active. This is supported by the lack of flares seen in a visual examination of the \tess\ SPOC simple aperture photometry (SAP) light curves. It is also consistent with the \cite{astudillo17} measurement of $\log R'_{HK} = -5.430 \pm 0.089$ based on HARPS spectra, which according to their study of the correlation between stellar rotation and $\log R'_{HK}$ corresponds to a rotation period of about 80 days (see \citealt{astudillo17} Equation~12).

To directly measure the stellar rotation period we used multi-season time-series photometry from WASP-South, described in \secr{wasp} and shown in \figr{wasp}, and concluded that the rotation period is $44\pm2$ days. That rotation period is consistent with a rotation of $51\pm14$ days measured by \cite{claytor23} using \tess\ data. The poorer precision of the latter is expected given the difficulties involved in measured a rotation period longer than the duration of a \tess\ sector with \tess\ data.

A rotation period of $44\pm2$ days is shorter than predicted by \cite{astudillo17} based on the spectroscopic $\log R'_{HK}$. However, their data show a large scatter around the relation between $\log R'_{HK}$ and rotation period (see their Figure 6), with one point, Gl 618A, with a mass of 0.38 \msun, having a measured rotation period of 56.5 days and a predicted rotation period based on their fitted relation of 78 days.

\section{Rejecting false positive scenarios}
\label{sec:fps}

In the subsections below we consider the various false positive scenarios, other than a transiting planet orbiting \target, that can lead to a transit-like signal in the \tess\ data, and show that they are rejected or found to be extremely unlikely, based on the data we have accumulated.

We first divided the \tess\ data into data subsets, and compared the transit light curve shape in different subsets. We used two 1-year long subsets, and four 6-months long subsets. In each subset the transit light curve shape was consistent with that of the other subsets.

\subsection{Is the M dwarf an eclipsing binary?}
\label{sec:eb}

In a common false positive scenario the transiting object is in fact a star, but smaller and cooler than the primary star in the system such that the secondary eclipse is not detected. To check this possibility we examine archival radial velocities (RVs) obtained by HARPS, as published by \cite{trifonov20}, shown in \figr{rvs} and listed in \tabr{rvs}. The available eight RVs have a span of 15 \ms, an RMS of 4.7 \ms, and their MAD $\times$ 1.48 (which equals the standard deviation for a normal distribution) is 2.7~\ms. While the RVs show a long-term RV variation on a time scale of several hundred days (see \figr{rvs} left panel), the RVs within a single observing season show a scatter of only a few \ms\ (see \figr{rvs} center panel). The five HARPS RVs taken during the second observing season in which this target was observed\footnote{Those five RVs were obtained between UT 2010 February 7 and UT 2010 March 28.} span 5.9 \ms, have an RMS of 3.5 \ms, and their MAD $\times$ 1.48 is 1.5 \ms. Clearly these RVs rule out the possibility of the transiting object being a star, including a stellar remnant, since a 0.08 \msun\ companion would result in a 22.4 \kms\ semi-amplitude assuming a circular orbit and given the orbital period and host star mass. We can also rule out the transiting companion being a massive planet since a Saturn mass companion ($\approx$0.3 \mjup) would result in a 90 \ms\ semi-amplitude.

Another way to rule out the transit companion being a star or a brown dwarf is by looking for orbital phase modulations in the \tess\ phase folded light curve, since a massive orbiting companion is expected to induce phase modulations along the orbit \citep[e.g.,][]{zucker07, shporer17,shporer19,wong20a,wong20b,wong21a,wong21b}. We tested the detectability of such a signal through injection and recovery of a sine wave signal centered on phase zero, which corresponds to the expected shape of a Doppler boosting signal \citep[e.g.,][]{loeb03, shporer10}. We ignore other potential phase modulations since they are expected to be much smaller in amplitude. When injecting a sinusoidal signal with a semi-amplitude of 9 ppm to the \tess\ data we recovered it with \sig{5} confidence. Therefore, larger semi-amplitudes can be detected in the \tess\ data. An amplitude of 9 ppm corresponds to a Doppler boosting signal induced by a 1.7 \mjup\ transiting companion, given the orbital period and the properties of the primary star. Therefore we conclude that more massive companions would have induced detectable phase variations. We have also attempted to fit a sine wave to the phase folded light curve without injection, while ignoring the in-transit data, and measured an amplitude of $0.3 \pm 1.8$ ppm, i.e., a null result.

Therefore, the above shows that if (1) the transit seen in \tess\ data originates from the target, and, if (2) the target is a single star, then the transiting companion must be an object smaller and less massive than a giant planet. The suppositions (1) and (2) are investigated in the following subsections.

\subsection{Does the transit-like signal originate from an unbounded eclipsing binary?}
\label{sec:beb}

The transit signal in the \tess\ data might originate from a stellar eclipsing binary that is not associated with the target but whose light is blended with the target in the \tess\ PSF, which is roughly an arcminute wide. Although, the SPOC pipeline difference images show that the transit signal is located $1.65 \pm 3.05$ arcsec from \target\ (see \secr{tess}). Yet, this angular distance can still allow for blending with other stars on a similar line of sight. To investigate this possibility we examined archival images and also obtained ground-based photometric time series during predicted transit times, with a typical angular resolution of one or a few arcseconds. 

We used archival images to look for stars at the current target location, that would be blended with the target in ground-based observations. We were able to do that thanks to the target's high proper motion, of $892.633\pm0.025$ mas per year. We used archival images of the target's location taken by the UK Schmidt Telescope (UKST) at 1979 and 1995 and made available by the STScI Digitized Sky Survey, shown in \figr{fcharts} along with a recent image from 2023 taken by the Las Cumbres Observatory 1m/Sinistro \citep{brown13}. The images show that the current target position does not contain any stars brighter than 19th mag that will be blended with the target in ground-based photometric observations. We used that brightness limit since fainter stars cannot explain the transit depth seen in \tess\ data even if they undergo a 100\% deep eclipse (full occultation), given the target's brightness and observed transit depth. We also checked that all stars brighter than 19th mag in the vicinity of the target, according to the TIC, are detectable in the archival images. The two stars closest to the target seen in \figr{fcharts} are at a distance of about 10 arcsec or more and are investigated below.

We note that it is in principle possible for a distant background star to be moving at a similar proper motion to that of the target so it is blended with it in all three epochs shown in \figr{fcharts}. However, that possibility is extremely unlikely. 

As described in \secr{lco}, we collected ground-based light curves of all stars close to the target and brighter than $T$ = 19.0 mag, and found that none of them shows a deep eclipse that could result in the shallow transit seen in \tess\ data due to blending with the target. This includes the two stars closest to the target marked in \figr{fcharts} and whose light curves are plotted in \figr{neb}. We have also confirmed that all stars brighter than $T$ = 19.0 mag in the vicinity of the target are detectable in the ground-based photometric data.

Therefore, we ruled out all stars detected in the data near the target as the source of an eclipse that can explain the transit seen in \tess\ data, and, determined that stars not detected in the data are too faint to be the origin of the transit even if they undergo a 100\% deep eclipse. We are left with the possibility that the transit seen in \tess\ data originates from the target star.

\subsection{Does the transit-like signal originate from a gravitationally bound eclipsing binary?}
\label{sec:beb}

The transit-like feature in \tess\ light curves might originate from an eclipsing binary that is gravitationally bound to the target star, so it has the same proper motion as the target and is not resolved in photometric and spectroscopic ground-based observations. 

To investigate this scenario we observed the target with high angular resolution imaging using the Gemini-South/Zorro speckle camera, at 562 nm and 832 nm, and using SOAR at the $I$-band, as described in \secr{imaging}. The contrast curves are shown in \figr{highangimage}. The 832 nm contrast curve places a \sig{5} upper limit showing no stars down to a magnitude difference of about 4.5 mag at 0.1 arcsec, 6.5 mag at 0.5 arcsec, and 7.0 mag at 1.0 arcsec. 
Given the properties of the target star, and using the tabulation provided by \cite{baraffe15} of stellar mass, radius, and effective temperature, a bounded main sequence star that is 4.5 mag fainter than the target at 832 nm has a mass of about 0.09 \msun.
The three angular distances listed above correspond to sky-projected distances of 1.5 AU, 7.6 AU, and 15.2 AU, respectively.

Further constraints on the existence of a stellar companion to the target come from the \gaia\ DR3 catalog \citep{gaia22}, although they are difficult to quantify. The \gaia\ DR3 catalog does not contain stars near the target down to a typical angular resolution of 0.4 arcsec and a completeness brightness beyond 20th magnitude \citep{riello21, gaia_dr3}. In addition the target's \gaia\ DR3 Re-normalised Unit Weight Error (RUWE) is 1.12 and the RV error is 0.17 \kms, which are typical for single stars \citep{katz19}.

Still, despite all the constraints above there could be a stellar companion to the target that is fainter than the target and at a sky-projected distance at the level of 1 AU, which corresponds to an angular distance of 0.066 arcsec given the distance to the target.
We turn to the \tess\ light curve shape to further constrain this scenario. Based on our transit fit, we find a \sig{3} upper bound on the ratio between the ingress and egress duration and that of the full transit, $t_{12} / t_{14}$, of 0.13. Since $R_p/R_s \lesssim t_{12}/t_{14}$, we can place a \sig{3} upper bound on the eclipsing (or transiting) object radius of $0.13 \times 0.43\, R_\odot = 6.1\, \rearth = 0.54\, R_J$. In other words, the shape of the transit light curve -- specifically the duration of the ingress/egress relative to full transit -- shows that the eclipsing/transiting object must be smaller than a star or brown dwarf \citep[e.g.,][]{chabrier00, sorahana13}. This is true whether the transited star is the known M dwarf or a companion star, since any companion star must be smaller in radius than the observed M dwarf. Therefore the transit seen in \tess\ data cannot be due to an eclipsing binary that is gravitationally bound to the target star. 

\subsection{Does the transiting planet orbit a gravitationally bound star?}
\label{sec:btr}

We are left with the possibility that a gravitationally bound star hosts the transiting planet. In this scenario the transiting object has negligible mass compared to its host star, so we can use our transit fit, and transit duration, to compute the host star's mean density. From the fit, we find $\rho = 5.1^{+1.4}_{-1.7}$~g~cm$^{-3}$ with a \sig{3} upper bound of $9.4$~g~cm$^{-3}$. This is consistent with the density of \target\ which is $7.64 \pm 0.32$~g~cm$^{-3}$, but inconsistent with the density of most main sequence stars cooler than \target\ since stellar density grows larger as stellar mass grows smaller. More quantitatively, stars smaller than $0.35\, M_\odot$ have mean densities larger than about 11.5~g~cm$^{-3}$. And if the transit host was a gravitationally bound star more massive than $0.35\, M_\odot$ then we would have detected its spectral lines in the HARPS spectra, since it would be roughly half as bright as the target star in the optical. Therefore, we can rule out a gravitationally bound star-planet system as the source of the \tess\ transit signal.

\subsection{Statistical validation}
\label{sec:statval}

To quantify the probability of the planet candidate being a false positive, we analyzed the \tess\ data using the statistical validation tool TRICERATOPS \citep{giacalone20, giacalone21} while accounting for follow-up data collected here. TRICERATOPS assesses the reliability of a planet candidate by simulating an array of astrophysical transit-producing scenarios, including: (1) a planet transiting the target star, (2) a stellar companion eclipsing the target star, (3) a planet transiting an unresolved star that is gravitationally bound to or chance aligned with the target star, (4) an unresolved pair of eclipsing binary stars that are gravitationally bound to or chance aligned with the target star, (5) a planet transiting a nearby star that is spatially resolved from the target star, or (6) a nearby eclipsing binary star that is spatially resolved from the target star. Using these simulations, it calculates the false positive probability (FPP) and nearby false positive probability (NFPP) for the planet candidate. The former takes into account all false positive scenarios (i.e., scenarios 2--6), whereas the latter only takes into account false positive scenarios originating from known nearby stars (i.e., scenarios 5--6). 

Because we fully rule out scenarios 5 and 6 with ground-based follow-up observations (see \secr{beb}), we set the NFPP to zero and calculate only the FPP. As an additional constraint, we fold in the 832 nm speckle contrast curve shown in \figr{highangimage}, which restricts the extent of the parameter space in which unresolved companions or chance-aligned stars can exist.
We ran TRICERATOPS 20 times, resulting in ${\rm FPP} = 3.80 \pm 0.80 \times 10^{-4}$, supporting the hypothesis that the target is a transiting star-planet system.

\section{Discussion}
\label{sec:dis}

\begin{figure*}
\begin{center}
\includegraphics[trim={0 0 0 0},clip,width=9.0cm]{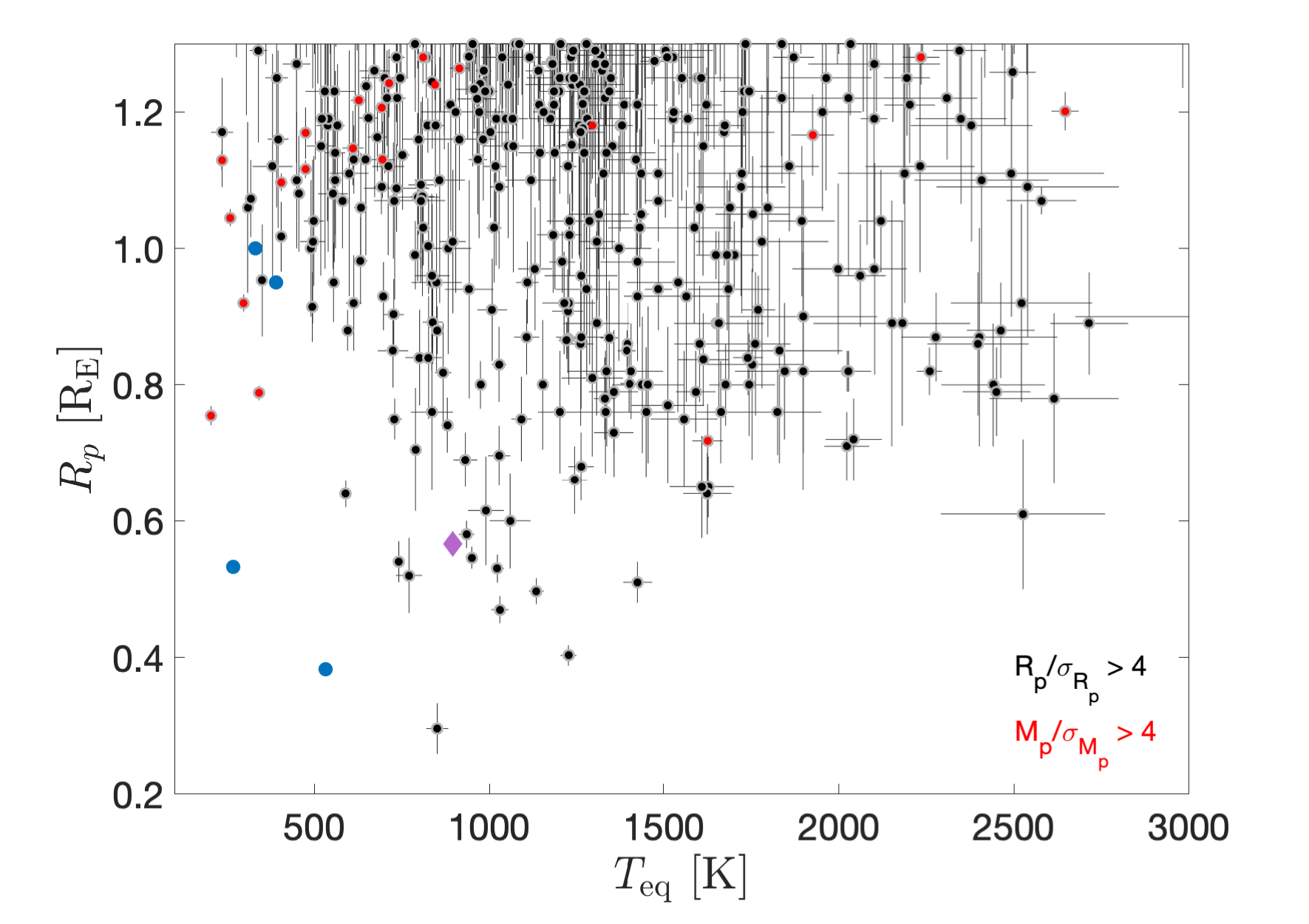}
\hspace{-4mm}
\includegraphics[trim={0 0 0 0},clip,width=9.0cm]{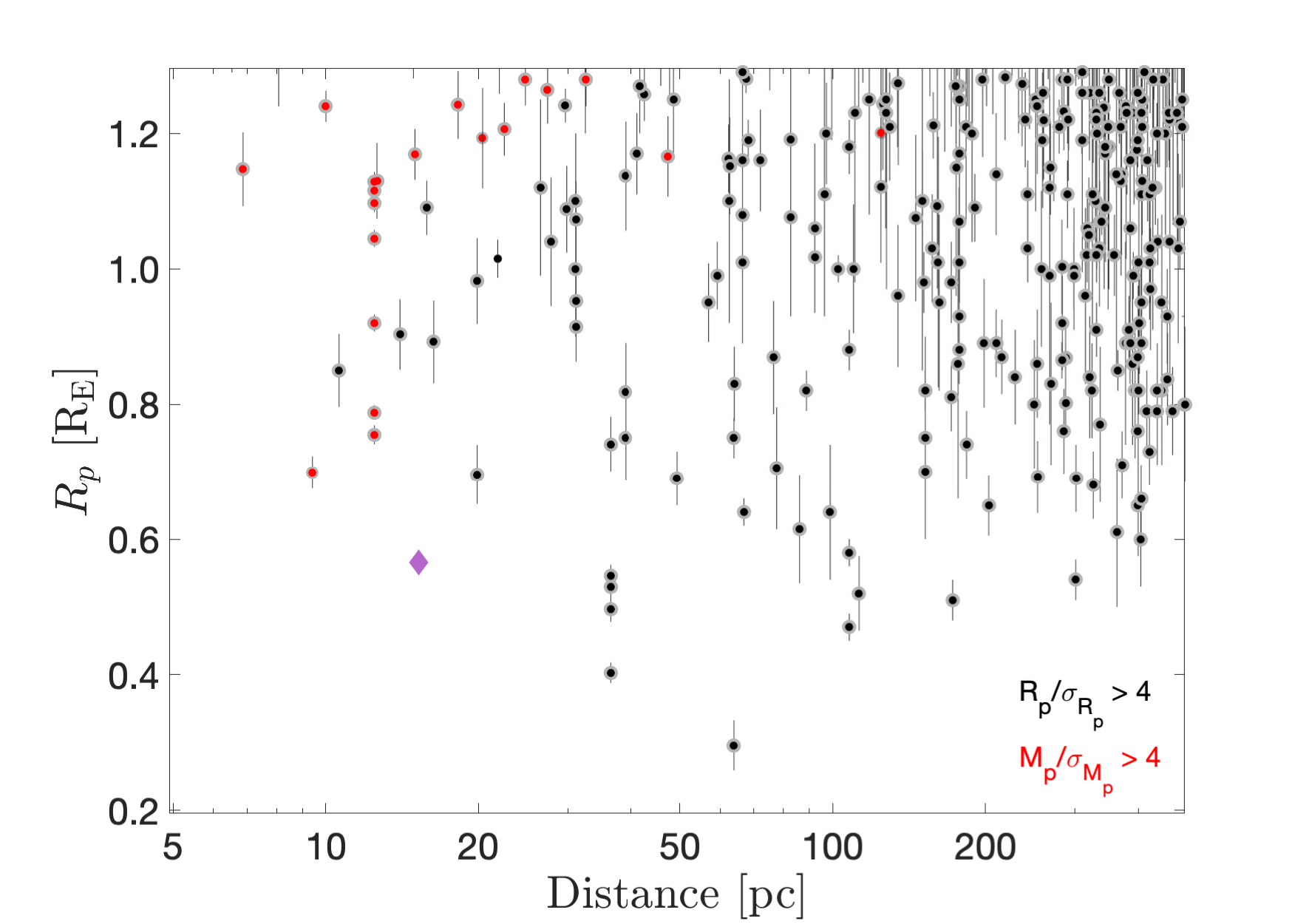}
\caption{{\bf Left:} Planet radius vs.~equilibrium temperature at the planet's orbit assuming zero bond albedo and no heat circulation. 
{\bf Right:} Planet radius vs.~host star distance to the Sun in log scale.
In both panels, planets with a radius measurement better than \sig{4} are marked in black, and those which also have their mass measured to better than \sig{4} are marked in red. \planet\ is marked by a purple diamond. Solar System terrestrial planets are marked in blue in the left panel. 
The figure is based on data from the NASA Exoplanets Archive downloaded on 2023 December 18.
}
\label{fig:radteqdist}
\end{center}
\end{figure*}

We present here the discovery and validation of \planet, with a radius of $0.566 \pm 0.014$ \rearth\ ($= 1.064 \pm 0.026$ Mars radius), making it one of the smallest known planets, and the smallest discovered by the \tess\ mission so far. \figr{radteqdist} and \figr{radmass} show \planet\ in the context of currently known exoplanets. \figr{radteqdist} left panel shows the radii of small planets for which the radius uncertainty is smaller than 25\%, as a function of equilibrium temperature (\teq). The equilibrium temperature was computed under the assumptions that the Bond albedo is zero and there is no heat circulation from the planet's dayside and nightside hemispheres, and the dayside has a uniform temperature. \figr{radteqdist} right panel shows the planet radius as a function of the host star's distance to the Sun, in log scale. In both panels planets with a mass measurement uncertainty smaller than 25\% are marked in red, and \planet\ is marked by a purple diamond. 

The plots show that there are only about a dozen known exoplanets with a directly measured radius smaller than or similar to that of \planet. All of them are validated exoplanets without a measured mass and most are in multi transiting planet systems \citep{muirhead12, barclay13, borucki13, marcy14, campante15, morton16, canas22}. It is interesting to note that most of these planets have \teq\ below 1{,}500~K, as seen in \figr{radteqdist} left panel. That is likely related to the fact that these planets are hosted mostly by K- and M-type stars, whose lower \teff\ and small radius compared to G- and F-type stars leads to a lower planet \teq\ for planets with similar orbital periods. And as these are the smallest known exoplanets with a directly measured radius, the fact that they orbit cool and small stars is expected.

\begin{figure}
\hspace{-8mm}
\includegraphics[width=10cm]{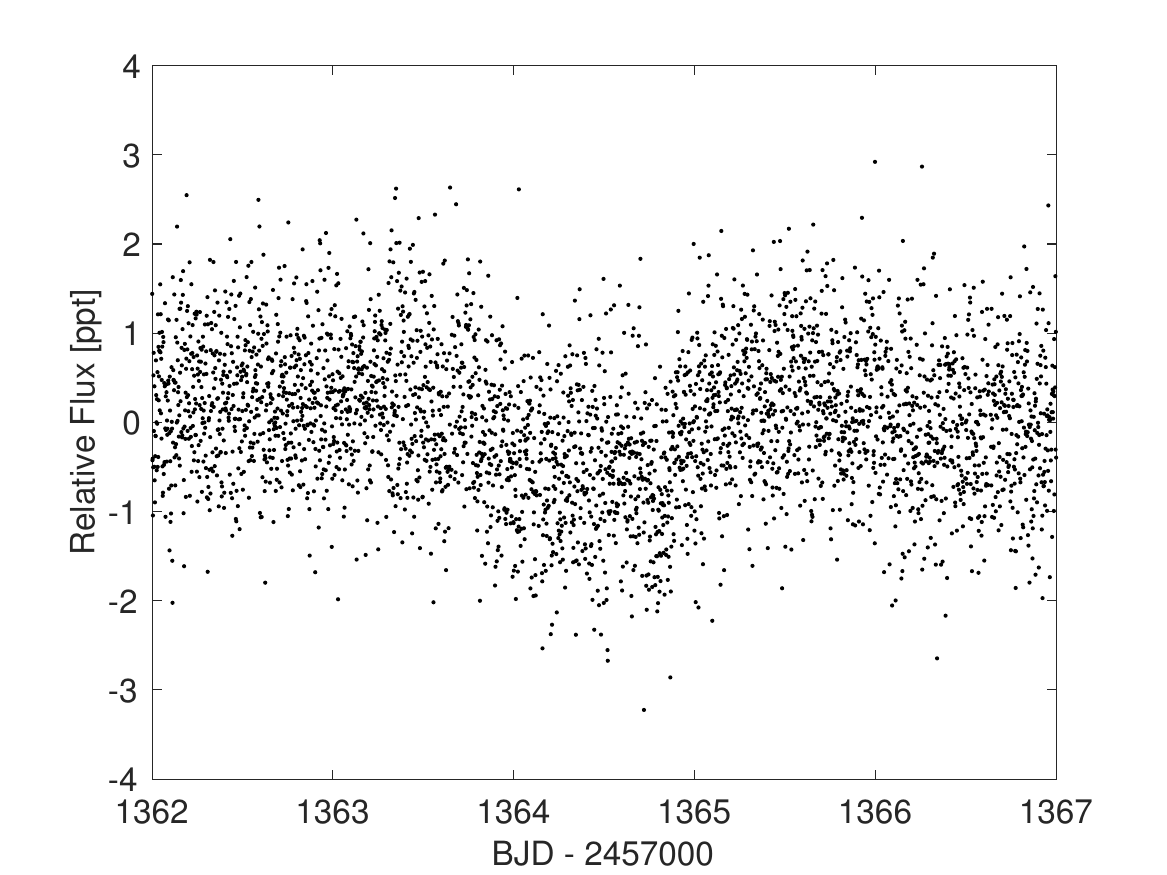}
\caption{A single transit-like feature that appears in the \tess\ light curve of \target\ during Sector 2. The plot shows relative flux, in parts per thousand (ppt), as a function of time in days (BJD). If this feature is due to a transiting planet then its depth of bout 0.6 ppt corresponds to a planet radius of about 3.6 \rearth, close to the size of Neptune. 
}
\label{fig:single}
\end{figure}

As mentioned above, most known exoplanets similar in size to \planet\ are in systems where there is at least one other known transiting exoplanet. But, despite the large amount \tess\ data collected for \target\ only one planet has been detected so far. It could be that the transit signals of other planets in the system are too small to be detectable in current data, and might be detectable with additional data to be collected by \tess\ in the future. It is also possible that other planets are not transiting, but might be detectable with precise radial velocity monitoring. It is interesting to note here that the SPOC analysis has identified a single transit-like signal close to BJD~=~2458364.6, during Sector 2, plotted in \figr{single}. The nature of that signal, whether it is due to a transiting planet, another astrophysical scenario (false positive), or a residual systematic feature, is still under investigation.

The validation of \planet\ motivates further study of the system, explored in the following subsections.

\subsection{Planet mass measurement}
\label{sec:mass}

\begin{figure}
\hspace{-8mm}
\includegraphics[width=10cm]{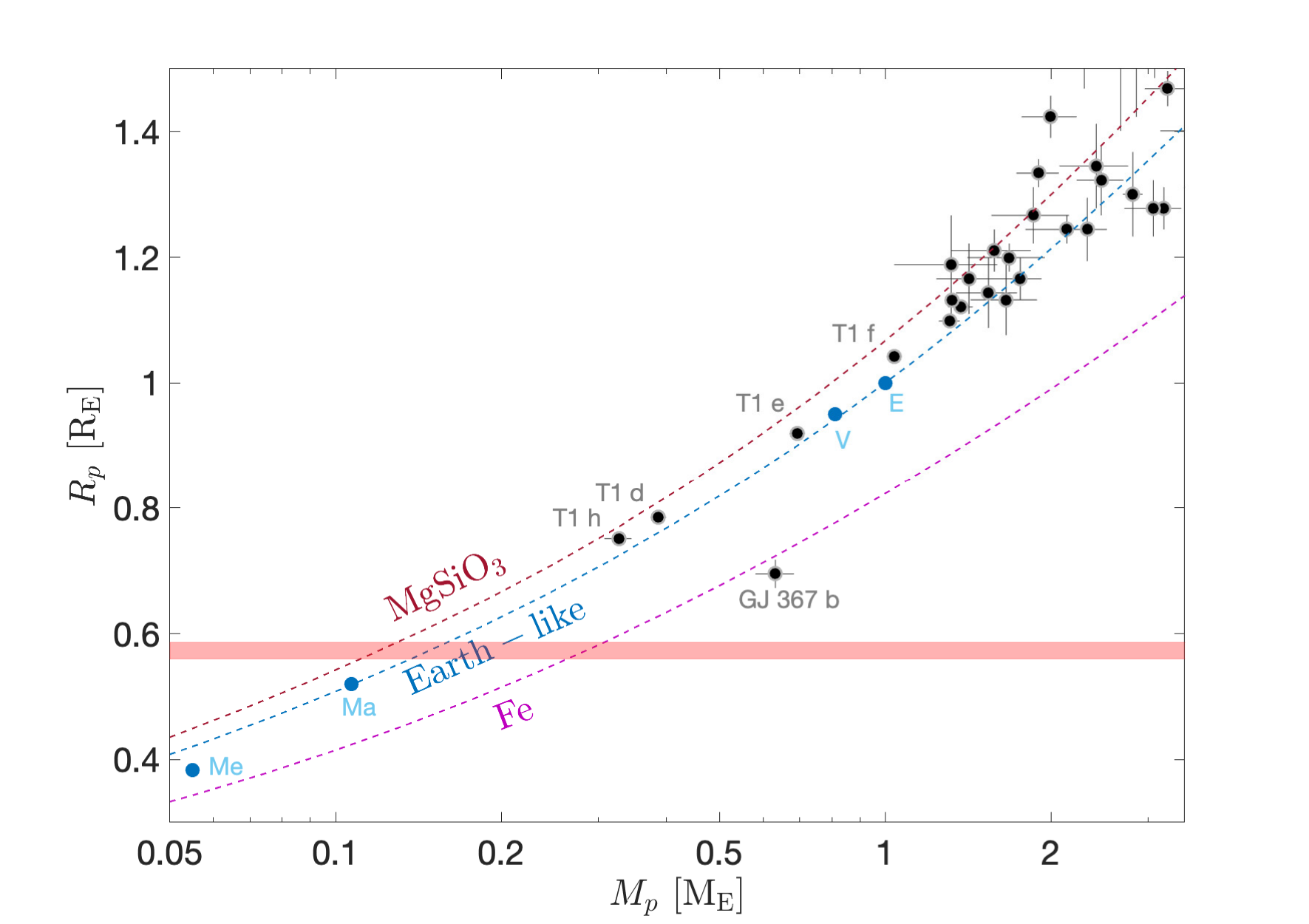}
\caption{Planet radius (linear scale) vs.~planet mass (log scale) for known transiting small planets with mass and radius measurement beyond a \sig{4} significance. \planet\ radius is marked by the red horizontal band whose width corresponds to the $\pm\sig{1}$ measured radius uncertainty. The dashed lines are composition models from \citet{zeng16}, including, from bottom to top, pure iron (100\% Fe; dashed purple line), Earth-like composition (32.5\% Fe + 67.5\% MgSiO$_3$; dashed blue line), and pure rock (100\% MgSiO$_3$; dashed brown line). A few of the planets are labeled where T1 is an abbreviation for TRAPPIST-1. 
Solar System terrestrial planets are marked in blue, where Earth is labeled by E, Venus by V, Mars by Ma, and Mercury by Me. 
The figure is based on data from the NASA Exoplanets Archive downloaded on 2023 December~18.
}
\label{fig:radmass}
\end{figure}

One natural next step is to attempt measuring the planet's mass. A planet mass measurement, combined with the known radius will provide the mean density and allow to constrain the planet's composition. \figr{radmass} shows the radius-mass diagram for small planets with measured radius and mass better than \sig{4}, with the radius of \planet\ marked by a horizontal red band. A planet mass measurement will allow comparisons to the Solar System's terrestrial planets, and test whether \planet\ is iron-rich like GJ~367~b \citep{lam21}.

\target\ shows low stellar activity (see \secr{star}), and the archival HARPS RVs show high RV stability, especially when examining RVs obtained within the same observing season (see \figr{rvs}). Therefore, this system seems to be suitable for intensive high precision RV measurements. Still, the host star's expected RV semi-amplitude is only $K_{RV}\approx$ 0.14 \ms\ for a planet mass of $M_p=0.15$ \mearth, which is the expected mass assuming Earth-like composition (see \figr{radmass}). We have carried out a simple simulation to estimate the number of RV measurements required for \planet\ mass measurement. In our simulation, we injected an RV circular orbit signal with $K_{RV}$=0.14 \ms\ into $N$ RVs, each with a precision of $\sigma_{RV}=0.25$ \ms, that are obtained once per night with a jitter of 1.5 hours in the measurement mid-exposure time. We adopted an RV uncertainty of 0.25 \ms\ as that is approximately the RV precision that can be obtained for stars similar to \target\ with state of the art high resolution Echelle spectrographs, e.g., ESPRESSO \citep{pepe14} and MAROON-X \citep[although it cannot observe \target\ position]{seifahrt18}.

We repeated the simulation $10^6$ times in order to obtain a large sample of measured $K_{RV}$, and took the mean and RMS of that distribution as the resulting $K_{RV}$ and its uncertainty. We repeated this exercise while changing the number of RVs ($N$), and found that $N$=155 RVs are needed for a \sig{5} measurement of the RV signal. While that is a large number, it is comparable to the number of RVs obtained in order to measure the mass of GJ~367~b \citep{lam21}.

The simulation described above is simplistic for several reasons. It assumes uniform RV precision that is not impacted by observing conditions, and that the target does not exhibit any other RV signals, due to, e.g., stellar activity and/or other planets in the system. Although, the star shows low activity and the stellar rotation is much longer than the orbital period (see \secr{star}). 
On the other hand, scheduling observations close to phases of quadrature, when the RV is at extrema (assuming a circular orbit) can lead to a higher RV amplitude measurement SNR. It is also possible that the planet's composition is more iron-rich than the Earth-like composition assumed in the simulation, in which case the number of RVs needed for a \sig{5} mass measurement is reduced, down to $N$=45 RVs for a pure iron composition where the planet mass is 0.28 \mearth.

\subsection{Planet atmosphere}
\label{sec:atm}

\begin{figure}
\includegraphics[width=8.2cm]{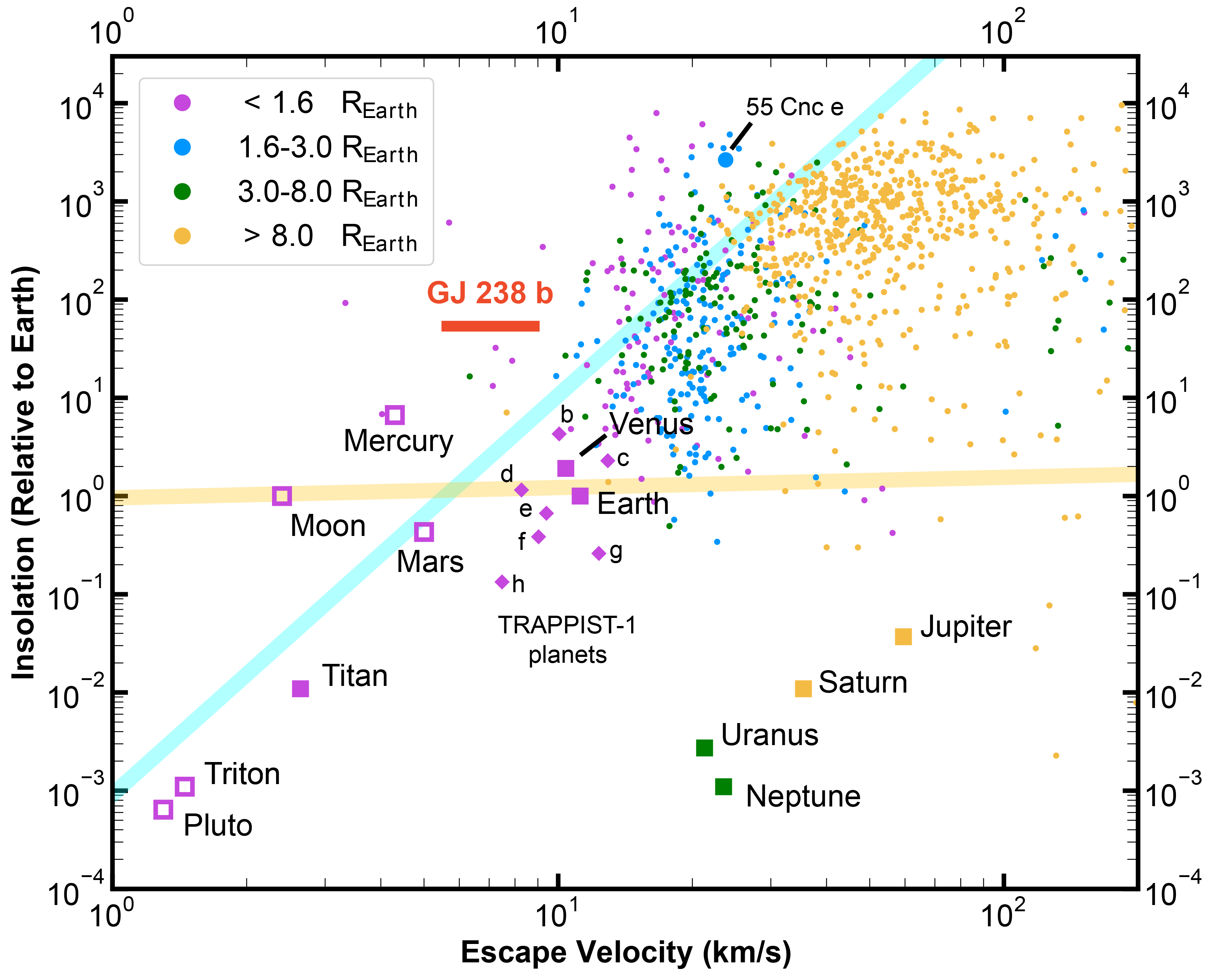}
\caption{Stellar insolation at a planet's orbit as a function of escape velocity from the planet's surface, in log-log scale.
Based on data from the NASA Exoplanet Archive downloaded on May 18, 2023. The empirical ``cosmic shoreline'' and the water vapor greenhouse runaway threshold \citep{ZahnleCatling17} are shown in cyan and yellow, respectively. Planets are categorized into terrestrial planets (magenta), sub-Neptunes (blue), Neptune-like planets (green), and gas giants (yellow) based on radius, see legend. The red line shows the possible range of escape velocities of GJ 238 b. Because only radius, but not mass, is known, the two endpoints represent two extreme interior scenarios -- a 100\% iron composition and a 100\% silicate composition.
}
\label{fig:insvel}
\end{figure}

Here we assess the prospect of studying the atmosphere of \planet. Empirically, small and highly irradiated bodies like \planet\ tend to lack substantial atmospheres. Atmospheric retention depends on the balance between insolation $I$ and a planet's ability to gravitationally bind its atmosphere, measured by escape velocity $v_{\rm esc}$. \cite{ZahnleCatling17} reported an empirical $I \propto v_{\rm esc}^4$ ``cosmic shoreline'' dividing between planets with and without an atmosphere, shown in \figr{insvel}. Given the small size of \planet, even under the densest pure-iron composition assumption it receives excessive insolation relative to its gravitational binding energy, placing \planet\ on the atmosphereless side of the empirical shoreline. M dwarfs have prolonged high-luminosity pre-main-sequence phase, implying that \planet\ would have been further above the shoreline before landing up in its current location.

However, planets above the shoreline are not necessary airless. Provided that abundant high-molecular-weight volatiles are stored in the mantle, survive from initial loss, and can degas with sufficiently high rate, revival of a secondary atmosphere is possible \citep[e.g.,][]{kite20_revival}. Such a high-molecular-weight atmosphere is resistant to escape. It has been suggested that the presence of a high-molecular-weight atmosphere on 55 Cnc e, a highly irradiated super-Earth above the ``cosmic shoreline'' (\figr{insvel}), is consistent with observations \citep[e.g.,][]{AngeloHu17, Jindal20}.

To investigate whether \planet\ can be detected in emission we modeled its emission spectrum assuming a blackbody with zero heat redistribution (no atmosphere), a 100 bar CO$_2$ atmosphere, and a 10 bar O$_2$ atmosphere using petitRADTRANS \citep{molliere19_prt}. The relative flux of the emission signal in these three scenarios is at most $\sim15$ ppm, which is below the 25 ppm systematic noise floor of the MIRI LRS instrument of JWST reported by \cite{Bouwman23_miri}. We have also investigated the possibility of detecting the atmosphere in transmission, and concluded that the signal is smaller than 10 ppm and hence undetectable by JWST. Therefore, we estimate that detecting the atmosphere of \planet, if it has any, is not feasible given current observational capacities.

\subsection{Astrometry}
\label{sec:ast}

At a distance of only 15.2 pc, \target\ is one the closest stars hosting a small terrestrial transiting planet, and the closest among planets of similar or smaller radius, as shown in \figr{radteqdist} right panel. This short distance will allow detection of planets on wide orbits with \gaia\ time series astrometry \citep{perryman14}, which will become available over the next few years \citep{gaia16, gaia22}. Given the distance to the system and the expected astrometric performance described in \cite{perryman14}, \gaia\ astrometry will be sensitive to a $\approx0.22\, \mjup$ planet at 1~AU, and a Neptune-mass planet at about 4~AU (although for the latter the \gaia\ astrometric time series may not cover a full orbital period).


\section{Summary}
\label{sec:sum}

We have shown that all false positive scenarios of the transiting planet candidate TOI-486.01, orbiting \target, are not viable or are extremely unlikely, making it a real planet, \planet\, orbiting the M dwarf \target. The ability to validate this planet relied critically on its host star's high proper motion and small radius and mass. We followed procedures applied to other similar systems \citep[e.g.,][]{vanderspek19, shporer20, gan20, tey23}.

The planet's radius of $0.566 \pm 0.014$ \rearth\ makes it one of the smallest planets currently known. Ongoing and future observations by \tess\ Extended Missions are expected to lead to detections of increasingly small transiting planets \citep{kunimoto22}, including terrestrial sub-Earth size planets such as \planet, which can be studied in further detail thanks to their proximity to the Sun and host star brightness. This will lead to gradually putting the solar system terrestrial planets in the context of exoplanets.

\acknowledgments

Funding for the \tess\ mission is provided by NASA's Science Mission directorate.

We acknowledge the use of public \tess\ data from pipelines at the \tess\ Science Office and at the \tess\ Science Processing Operations Center. 

This paper includes data collected by the \tess\ mission, which are publicly available from the Mikulski Archive for Space Telescopes (MAST).

Resources supporting this work were provided by the NASA High-End Computing (HEC) Program through the NASA Advanced Supercomputing (NAS) Division at Ames Research Center. 

This research has made use of the NASA Exoplanet Archive, which is operated by the California Institute of Technology, under contract with NASA under the Exoplanet Exploration Program.

The Digitized Sky Surveys were produced at the Space Telescope Science Institute under U.S.~Government grant NAG W-2166. The images of these surveys are based on photographic data obtained using the Oschin Schmidt Telescope on Palomar Mountain and the UK Schmidt Telescope. The plates were processed into the present compressed digital form with the permission of these institutions.

This work makes use of observations from the LCOGT network. Part of the LCOGT telescope time was granted by NOIRLab through the Mid-Scale Innovations Program (MSIP). MSIP is funded by NSF.

Based in part on observations obtained at the Southern Astrophysical Research (SOAR) telescope, which is a joint project of the Minist\'{e}rio da Ci\^{e}ncia, Tecnologia e Inova\c{c}\~{o}es (MCTI/LNA) do Brasil, the US National Science Foundation’s NOIRLab, the University of North Carolina at Chapel Hill (UNC), and Michigan State University (MSU).

Some of the observations in this paper made use of the High-Resolution Imaging instrument Zorro and were obtained under Gemini LLP Proposal Number: GN/S-2021A-LP-105. Zorro was funded by the NASA Exoplanet Exploration Program and built at the NASA Ames Research Center by Steve B. Howell, Nic Scott, Elliott P. Horch, and Emmett Quigley. Zorro was mounted on the Gemini South telescope of the international Gemini Observatory, a program of NSF’s OIR Lab, which is managed by the Association of Universities for Research in Astronomy (AURA) under a cooperative agreement with the National Science Foundation. on behalf of the Gemini partnership: the National Science Foundation (United States), National Research Council (Canada), Agencia Nacional de Investigación y Desarrollo (Chile), Ministerio de Ciencia, Tecnología e Innovación (Argentina), Ministério da Ciência, Tecnologia, Inovações e Comunicações (Brazil), and Korea Astronomy and Space Science Institute (Republic of Korea).

This work has made use of data from the European Space Agency (ESA) mission
{\it Gaia} (\url{https://www.cosmos.esa.int/gaia}), processed by the {\it Gaia}
Data Processing and Analysis Consortium (DPAC,
\url{https://www.cosmos.esa.int/web/gaia/dpac/consortium}). Funding for the DPAC
has been provided by national institutions, in particular the institutions
participating in the {\it Gaia} Multilateral Agreement.

Z.L.~acknowledges funding from the Center for Matter at Atomic Pressures (CMAP), a National Science Foundation (NSF) Physics Frontiers Center, under Award PHY-2020249.

KAC acknowledges support from the \tess\ mission via subaward s3449 from MIT.\\

SG is supported by an NSF Astronomy and Astrophysics Postdoctoral Fellowship under award AST-2303922.

\facilities{ESO:3.6m (HARPS), Exoplanet Archive, Gemini South 8-m (Zorro), LCO:1m (Sinistro), SOAR: 4.1m (HRCam), \tess}
\software{
AstroImageJ \citep{collins17},
TAPIR \citep{jensen13}
}




\begin{thebibliography}{}







\bibitem[Angelo \& Hu(2017)]{AngeloHu17} Angelo, I., Hu, R.\ 2017, \aj, 154, 232. doi:10.3847/1538-3881/aa9278

\bibitem[Astudillo-Defru et al.(2017)]{astudillo17} Astudillo-Defru, N., Delfosse, X., Bonfils, X., et al.\ 2017, \aap, 600, A13


\bibitem[Barclay et al.(2013)]{barclay13} Barclay, T., Rowe, J.~F., Lissauer, J.~J., et al.\ 2013, \nat, 494, 452. doi:10.1038/nature11914





\bibitem[Baraffe et al.(2015)]{baraffe15} Baraffe, I., Homeier, D., Allard, F., et al.\ 2015, \aap, 577, A42. doi:10.1051/0004-6361/201425481


\bibitem[Borucki et al.(2013)]{borucki13} Borucki, W.~J., Agol, E., Fressin, F., et al.\ 2013, Science, 340, 587. doi:10.1126/science.1234702

\bibitem[Bouwman et al.(2023)]{Bouwman23_miri} Bouwman, J. et al. 2023, \pasp, 135, 1045

\bibitem[Boyajian et al.(2012)]{boyajian12} Boyajian, T.~S., von Braun, K., van Belle, G., et al.\ 2012, \apj, 757, 112

\bibitem[Brown et al.(2013)]{brown13} Brown, T.~M., Baliber, N., Bianco, F.~B., et al.\ 2013, \pasp, 125, 1031


\bibitem[Campante et al.(2015)]{campante15} Campante, T.~L., Barclay, T., Swift, J.~J., et al.\ 2015, \apj, 799, 170. doi:10.1088/0004-637X/799/2/170

\bibitem[Ca{\~n}as et al.(2022)]{canas22} Ca{\~n}as, C.~I., Mahadevan, S., Cochran, W.~D., et al.\ 2022, \aj, 163, 3. doi:10.3847/1538-3881/ac3088


\bibitem[Chabrier \& Baraffe(2000)]{chabrier00} Chabrier, G. \& Baraffe, I.\ 2000, \araa, 38, 337. doi:10.1146/annurev.astro.38.1.337



\bibitem[Ciardi et al.(2015)]{ciardi15} Ciardi, D.~R., Beichman, C.~A., Horch, E.~P., et al.\ 2015, \apj, 805, 16





\bibitem[Claytor et al.(2023)]{claytor23} Claytor, Z.~R., van Saders, J.~L., Cao, L., et al.\ 2023, arXiv:2307.05664. doi:10.48550/arXiv.2307.05664


\bibitem[Collins et al.(2017)]{collins17} Collins, K.~A., Kielkopf, J.~F., Stassun, K.~G., et al.\ 2017, \aj, 153, 77



\bibitem[Cutri et al.(2003)]{cutri03} Cutri, R.~M., Skrutskie, M.~F., van Dyk, S., et al.\ 2003, VizieR Online Data Catalog, II/246





\bibitem[Dittmann et al.(2016)]{dittmann16} Dittmann, J.~A., Irwin, J.~M., Charbonneau, D., et al.\ 2016, \apj, 818, 153





\bibitem[Foreman-Mackey et~al.(2013)]{foreman13}
Foreman-Mackey, D., Hogg, D.~W., Lang, D., \& Goodman, J. 2013, \pasp, 125, 306


\bibitem[Foreman-Mackey et al.(2021)]{foreman21} Foreman-Mackey, D., Luger, R., Agol, E., et al.\ 2021, The Journal of Open Source Software, 6, 3285. doi:10.21105/joss.03285





\bibitem[Gaia Collaboration et al.(2016)]{gaia16} Gaia Collaboration, Prusti, T., de Bruijne, J.~H.~J., et al.\ 2016, \aap, 595, A1. doi:10.1051/0004-6361/201629272 

\bibitem[Gaia Collaboration et al.(2022)]{gaia22} Gaia Collaboration, Vallenari, A., Brown, A.~G.~A., et al.\ 2022, arXiv:2208.00211 



\bibitem[Gaia Collaboration et al.(2023)]{gaia_dr3} Gaia Collaboration, Vallenari, A., Brown, A.~G.~A., et al.\ 2023, \aap, 674, A1. doi:10.1051/0004-6361/202243940

\bibitem[Gan et al.(2020)]{gan20} Gan, T., Shporer, A., Livingston, J., et al.\ 2020, AAS Journals, submitted

\bibitem[Gelman \& Rubin(1992)]{gelman92} Gelman, A. \& Rubin, D.~B.\ 1992, Statistical Science, 7, 457. doi:10.1214/ss/1177011136

\bibitem[Giacalone \& Dressing(2020)]{giacalone20} Giacalone, S. \& Dressing, C.~D.\ 2020, Astrophysics Source Code Library. ascl:2002.004

\bibitem[Giacalone et al.(2021)]{giacalone21} Giacalone, S., Dressing, C.~D., Jensen, E.~L.~N., et al.\ 2021, \aj, 161, 24. doi:10.3847/1538-3881/abc6af









\bibitem[Huang et al.(2020a)]{huang20a} Huang, C.~X., Vanderburg, A., P{\'a}l, A., et al.\ 2020a, Research Notes of the American Astronomical Society, 4, 204. doi:10.3847/2515-5172/abca2e

\bibitem[Huang et al.(2020b)]{huang20b} Huang, C.~X., Vanderburg, A., P{\'a}l, A., et al.\ 2020b, Research Notes of the American Astronomical Society, 4, 206. doi:10.3847/2515-5172/abca2d

\bibitem[Husser et al.(2013)]{husser13} Husser, T.-O., Wende-von Berg, S., Dreizler, S., et al.\ 2013, \aap, 553, A6. doi:10.1051/0004-6361/201219058






\bibitem[Jenkins et~al.(2016)]{jenkins16} Jenkins, J.~M., Twicken, J.~D., McCauliff, S., et~al.\ 2016, \procspie, 9913, 99133E


\bibitem[Jensen(2013)]{jensen13} Jensen, E.\ 2013, Tapir: A web interface for transit/eclipse observability, ascl:1306.007

\bibitem[Jindal et~al.(2020)]{Jindal20} Jindal, A., de Mooij, E., Jayawardhana, R., et~al.\ 2020, \aj, 160, 101, doi:10.3847/1538-3881/aba1eb




\bibitem[Katz et al.(2019)]{katz19} Katz, D., Sartoretti, P., Cropper, M., et al.\ 2019, \aap, 622, A205


\bibitem[Kipping(2013)]{kipping13} Kipping, D.~M.\ 2013, \mnras, 435, 2152

\bibitem[Kite \& Barnett(2020)]{kite20_revival} Kite, E., Barnett, M. 2020, PNAS, 117, 31



\bibitem[Kunimoto et al.(2022)]{kunimoto22} Kunimoto, M., Winn, J., Ricker, G.~R., et al.\ 2022, \aj, 163, 290. doi:10.3847/1538-3881/ac68e3

\bibitem[Kuznetsov et al.(2019)]{kuznetsov19} Kuznetsov, M.~K., del Burgo, C., Pavlenko, Y.~V., et al.\ 2019, \apj, 878, 134. doi:10.3847/1538-4357/ab1fe9



\bibitem[Lam et al.(2021)]{lam21} Lam, K.~W.~F., Csizmadia, S., Astudillo-Defru, N., et al.\ 2021, Science, 374, 1271. doi:10.1126/science.aay3253






\bibitem[Loeb \& Gaudi(2003)]{loeb03} Loeb, A. \& Gaudi, B.~S.\ 2003, \apjl, 588, L117. doi:10.1086/375551






\bibitem[Mann et al.(2015)]{mann15} Mann, A.~W., Feiden, G.~A., Gaidos, E., et al.\ 2015, \apj, 804, 64

\bibitem[Mann et al.(2019)]{mann19} Mann, A.~W., Dupuy, T., Kraus, A.~L., et al.\ 2019, \apj, 871, 63

\bibitem[Maxted et al.(2011)]{maxted11} Maxted, P.~F.~L., Anderson, D.~R., Collier Cameron, A., et al.\ 2011, \pasp, 123, 547


\bibitem[Marcy et al.(2014)]{marcy14} Marcy, G.~W., Isaacson, H., Howard, A.~W., et al.\ 2014, \apjs, 210, 20. doi:10.1088/0067-0049/210/2/20


\bibitem[Mayor et al.(2003)]{mayor03} Mayor, M., Pepe, F., Queloz, D., et al.\ 2003, The Messenger, 114, 20

\bibitem[McQuillan et al.(2014)]{Mcquillan14} McQuillan, A., Mazeh, T., \& Aigrain, S.\ 2014, \apjs, 211, 24. doi:10.1088/0067-0049/211/2/24

\bibitem[Mollière et al.(2019)]{molliere19_prt} Mollière, P. et al. 2019, \aap, 627, A67

\bibitem[Morton(2015)]{morton15} Morton, T.~D.\ 2015, Astrophysics Source Code Library. ascl:1503.011

\bibitem[Morton et al.(2016)]{morton16} Morton, T.~D., Bryson, S.~T., Coughlin, J.~L., et al.\ 2016, \apj, 822, 86. doi:10.3847/0004-637X/822/2/86

\bibitem[Muirhead et al.(2012)]{muirhead12} Muirhead, P.~S., Johnson, J.~A., Apps, K., et al.\ 2012, \apj, 747, 144. doi:10.1088/0004-637X/747/2/144

\bibitem[Muirhead et al.(2018)]{muirhead18} Muirhead, P.~S., Dressing, C.~D., Mann, A.~W., et al.\ 2018, \aj, 155, 180












\bibitem[Pecaut \& Mamajek(2013)]{pecaut13} Pecaut, M.~J. \& Mamajek, E.~E.\ 2013, \apjs, 208, 9. doi:10.1088/0067-0049/208/1/9

\bibitem[Pepe et al.(2002)]{pepe02} Pepe, F., Mayor, M., Rupprecht, G., et al.\ 2002, The Messenger, 110, 9

\bibitem[Pepe et al.(2014)]{pepe14} Pepe, F., Molaro, P., Cristiani, S., et al.\ 2014, Astronomische Nachrichten, 335, 8. doi:12.1002/asna.201312004

\bibitem[Perryman et al.(2014)]{perryman14} Perryman, M., Hartman, J., Bakos, G. {\'A}., et al.\ 2014, \apj, 797, 14

\bibitem[Pollacco et al.(2006)]{pollacco06} Pollacco, D.~L., Skillen, I., Collier Cameron, A., et al.\ 2006, \pasp, 118, 1407




\bibitem[Ricker et al.(2014)]{ricker14} Ricker, G.~R., Winn, J.~N., Vanderspek, R., et al.\ 2014, \procspie, 914320


\bibitem[Riello et al.(2021)]{riello21} Riello, M., De Angeli, F., Evans, D.~W., et al.\ 2021, \aap, 649, A3. doi:10.1051/0004-6361/202039587






\bibitem[Scott et al.(2021)]{scott21} Scott, N.~J., Howell, S.~B., Gnilka, C.~L., et al.\ 2021, Frontiers in Astronomy and Space Sciences, 8, 138. doi:10.3389/fspas.2021.716560

\bibitem[Seifahrt et al.(2018)]{seifahrt18} Seifahrt, A., St{\"u}rmer, J., Bean, J.~L., et al.\ 2018, \procspie, 10702, 107026D. doi:10.1117/12.2312936

\bibitem[Shporer et al.(2010)]{shporer10} Shporer, A., Kaplan, D.~L., Steinfadt, J.~D.~R., et al.\ 2010, \apjl, 725, L200. doi:10.1088/2041-8205/725/2/L200

\bibitem[Shporer(2017)]{shporer17} Shporer, A.\ 2017, \pasp, 129, 072001 

\bibitem[Shporer et al.(2019)]{shporer19} Shporer, A., Wong, I., Huang, C.~X., et al.\ 2019, \aj, 157, 178

\bibitem[Shporer et al.(2020)]{shporer20} Shporer, A., Collins, K.~A., Astudillo-Defru, N., et al.\ 2020, \apjl, 890, L7. doi:10.3847/2041-8213/ab7020


\bibitem[Smith et~al.(2012)]{smith12} Smith, J.~C., Stumpe, M.~C., Van Cleve, J.~E., et~al. 2012, \pasp, 124, 1000


\bibitem[Sorahana et al.(2013)]{sorahana13} Sorahana, S., Yamamura, I., \& Murakami, H.\ 2013, \apj, 767, 77. doi:10.1088/0004-637X/767/1/77


\bibitem[Stassun \& Torres(2016)]{stassun16} Stassun, K.~G. \& Torres, G.\ 2016, \aj, 152, 180. doi:10.3847/0004-6256/152/6/180

\bibitem[Stassun et~al.(2017)]{stassun17} Stassun, K.~G., Collins, K.~A., \& Gaudi, B.~S.\ 2017, \aj, 153, 136 

\bibitem[Stassun et~al.(2018)]{stassun18a} Stassun, K.~G., Oelkers, R.~J., Pepper, J., et~al.\ 2018a, \aj, 156, 102 

\bibitem[Stassun et al.(2018)]{stassun18b} Stassun, K.~G., Corsaro, E., Pepper, J.~A., et al.\ 2018b, \aj, 155, 22. doi:10.3847/1538-3881/aa998a


\bibitem[Stumpe et al.(2012)]{stumpe12} Stumpe, M.~C., Smith, J.~C., Van Cleve, J.~E., et al.\ 2012, \pasp, 124, 985. doi:10.1086/667698

\bibitem[Stumpe et~al.(2014)]{stumpe14} Stumpe, M.~C., Smith, J.~C., Catanzarite, J. H., et~al. 2014, \pasp, 126, 100



\bibitem[Tayar et al.(2022)]{tayar22} Tayar, J., Claytor, Z.~R., Huber, D., et al.\ 2022, \apj, 927, 31. doi:10.3847/1538-4357/ac4bbc

\bibitem[Tey et al.(2023)]{tey23} Tey, E., Huang, C.~X., Kunimoto, M., et al.\ 2023, \aj, 165, 93. doi:10.3847/1538-3881/acaf88


\bibitem[Torres et al.(2011)]{torres11} Torres, G., Fressin, F., Batalha, N.~M., et al.\ 2011, \apj, 727, 24. doi:10.1088/0004-637X/727/1/24

\bibitem[Tokovinin(2018)]{tokovinin18} Tokovinin, A.\ 2018, \pasp, 130, 035002. doi:10.1088/1538-3873/aaa7d9

\bibitem[Trifonov et al.(2020)]{trifonov20} Trifonov, T., Tal-Or, L., Zechmeister, M., et al.\ 2020, \aap, 636, A74. doi:10.1051/0004-6361/201936686

\bibitem[Twicken et al.(2018)]{twicken18} Twicken, J.~D., Catanzarite, J.~H., Clarke, B.~D., et al.\ 2018, \pasp, 130, 064502. doi:10.1088/1538-3873/aab694




\bibitem[Vanderspek et al.(2019)]{vanderspek19} Vanderspek, R., Huang, C.~X., Vanderburg, A., et al.\ 2019, \apjl, 871, L24 


\bibitem[Walkowicz, \& Hawley(2009)]{walkowicz09} Walkowicz, L.~M., \& Hawley, S.~L.\ 2009, \aj, 137, 3297






\bibitem[Wong et al.(2020a)]{wong20a} Wong, I., Shporer, A., Kitzmann, D., et al.\ 2020a, \aj, 160, 88. doi:10.3847/1538-3881/aba2cb

\bibitem[Wong et al.(2020b)]{wong20b} Wong, I., Shporer, A., Daylan, T., et al.\ 2020b, \aj, 160, 155. doi:10.3847/1538-3881/ababad

\bibitem[Wong et al.(2021a)]{wong21a} Wong, I., Kitzmann, D., Shporer, A., et al.\ 2021a, \aj, 162, 127. doi:10.3847/1538-3881/ac0c7d

\bibitem[Wong et al.(2021b)]{wong21b} Wong, I., Shporer, A., Zhou, G., et al.\ 2021b, \aj, 162, 256. doi:10.3847/1538-3881/ac26bd





\bibitem[Zahnle \& Catling(2017)]{ZahnleCatling17} Zahnle, K., Catling, D.\ 2017, \apj, 843, 122. doi:10.3847/1538-4357/aa7846

\bibitem[Zacharias et al.(2012)]{UCAC4} Zacharias, N., Finch, C.~T., Girard, T.~M., et al.\ 2012, VizieR Online Data Catalog, I/322A

\bibitem[Zechmeister et al.(2018)]{zechmeister18} Zechmeister, M., Reiners, A., Amado, P.~J., et al.\ 2018, \aap, 609, A12. doi:10.1051/0004-6361/201731483

\bibitem[Ziegler et al.(2021)]{ziegler21} Ziegler, C., Tokovinin, A., Latiolais, M., et al.\ 2021, \aj, 162, 192. doi:10.3847/1538-3881/ac17f6

\bibitem[Zeng et al.(2016)]{zeng16} Zeng, L., Sasselov, D.~D., \& Jacobsen, S.~B.\ 2016, \apj, 819, 127


\bibitem[Zucker et al.(2007)]{zucker07} Zucker, S., Mazeh, T., \& Alexander, T.\ 2007, \apj, 670, 1326. doi:10.1086/521389


\end{thebibliography}
\end{document}